\documentclass[aps,12pt,superscriptaddress,floatfix]{revtex4} 
\usepackage[T1]{fontenc} 
\usepackage{lmodern}
\usepackage{hyperref}
\usepackage{graphicx}
\usepackage{bm}
\usepackage{amsmath}
\usepackage{amssymb}
\usepackage{dcolumn}
\usepackage{multirow}
\usepackage{color}
\usepackage{aas_macros}
\usepackage{fancyhdr}
\usepackage{tablefootnote}
\usepackage[papersize={8.5in,11in},margin=1.5cm]{geometry}

\begin{document}

\title{Neutrinos from type Ia supernovae: the deflagration-to-detonation transition scenario}

\author{Warren P. \surname{Wright}}
\email{wpwright@ncsu.edu}
\affiliation{Department of Physics,  North Carolina State University, Raleigh, North Carolina 27695, USA}

\author{Gautam \surname{Nagaraj}}
\email{grnagara@ncsu.edu}
\affiliation{Department of Physics,  North Carolina State University, Raleigh, North Carolina 27695, USA}

\author{James P. \surname{Kneller}}
\email{jpknelle@ncsu.edu}
\affiliation{Department of Physics,  North Carolina State University, Raleigh, North Carolina 27695, USA}

\author{Kate \surname{Scholberg}}
\email{schol@phy.duke.edu}
\affiliation{Department of Physics, Duke University, Durham, North Carolina 27708, USA}

\author{Ivo R. \surname{Seitenzahl}}
\email{ivo.seitenzahl@anu.edu.au}
\affiliation{Research School of Astronomy and Astrophysics, Australian National University, Canberra, Australia Capital Territory 2611, Australia} 
\affiliation{ARC Centre of Excellence for All-Sky Astrophysics (CAASTRO) 
}

\date{\today}
\begin{abstract}
It has long been recognized that the neutrinos detected from the next core-collapse supernova in the Galaxy have the potential to reveal important information about the dynamics of the explosion and the nucleosynthesis conditions as well as allowing us to probe the properties of the neutrino itself. The neutrinos emitted from thermonuclear - type Ia - supernovae also possess the same potential, although these supernovae are dimmer neutrino sources. For the first time, we calculate the time, energy, line of sight, and neutrino-flavor-dependent features of the neutrino signal expected from a three-dimensional delayed-detonation explosion simulation, where a deflagration-to-detonation transition triggers the complete disruption of a near-Chandrasekhar mass carbon-oxygen white dwarf. We also calculate the neutrino flavor evolution along eight lines of sight through the simulation as a function of time and energy using an exact three-flavor transformation code. We identify a characteristic spectral peak at $\sim 10$ MeV as a signature of electron captures on copper. This peak is a potentially distinguishing feature of explosion models since it reflects the nucleosynthesis conditions early in the explosion. We simulate the event rates in the Super-K, Hyper-K, JUNO, and DUNE neutrino detectors with the SNOwGLoBES event rate calculation software and also compute the IceCube signal. Hyper-K will be able to detect neutrinos from our model out to a distance of $\sim 10$ kpc. At 1 kpc, JUNO, Super-K, and DUNE would register a few events while IceCube and Hyper-K would register several tens of events.
\end{abstract}
\maketitle

\section{Introduction \label{Introduction}}
Type Ia supernovae (SNe Ia) hold a special place in our understanding of the Universe. SNe Ia act as standard candles \cite{Phillips1993,Phillips1999} for astronomical distance measurements. Most famously, this quality of SNe Ia as distance indicators was used to show that our Universe is expanding at an accelerating rate \cite{Riess1998,Schmidt1998,Perlmutter1999}.  Despite their importance, little is conclusively known about SN Ia progenitors and their explosion mechanism. The standard theory is that a SN Ia is a thermonuclear explosion of a white dwarf (WD) that gained enough mass to trigger explosive carbon burning. The mass gain mechanism is usually thought to be through interaction with a binary companion, although whether that system is single or double degenerate (or some other variant) remains unclear (see \cite{Maoz2014}\&\cite{Ruiz-Lapuente2014} for reviews). Regarding the explosion mechanism, many candidates have been studied. From the first pure detonation \cite{Arnett1969} and pure deflagration \cite{Nomoto1976} models to a plethora of more modern models including the delayed-detonation transition model (DDT) model, the gravitationally confined detonation model, the pulsational reverse detonation model, and many others (for a recent review see \cite{Hillebrandt2013} and references therein).

One reason why the progenitor problem remains unsettled is that, unlike for core-collapse SNe, no progenitor or companion stars have been identified in archival pre-explosion images, with one exception. The exception is the identification in archival Hubble Space Telescope images of the likely companion star of SN 2012Z \cite{McCully2014}. However, SN~2012Z was not spectroscopically normal and belongs to a faint sub-class of SNe Ia, the so-called 2002cx-like (or also Iax) SNe. To date, no pre-explosion identification of either the progenitor or the companion star for ``normal'' SNe Ia has been successful.

Attempts to pin-point the progenitor system based on the predicted spectral time evolution of the optical emission are often inconclusive \cite{Roepke2012}. The same holds for the inverse process of reconstructing the composition from the observed spectral evolution, i.e.,\ abundance tomography \cite{stehle2005,mazzali2008,hachinger2009,hachinger2013}. Efforts that compare the observed SN rate to predictions of the hypothesized formation channels from binary population synthesis calculations \cite{yungelson2000a,ruiter2009a,ruiter2011a,toonen2012a,ruiter2013a,mennekens2010a,wang2012a} are also inconclusive. 
Numerous other approaches that aim to identify the progenitor systems via more or less compelling observable signatures exist, including searching for the signature of the shocked companion \cite{kasen2010,bloom2012,marion2015}, time-variable Na absorption features \cite{patat2007,sternberg2011,dilday2012}, late-time bolometric light curves \cite{seitenzahl2009d,kerzendorf2014,graur2016}, gamma-ray emission \cite{sim2008,maeda2012,summa2013}, the chemical composition of supernova remnants \cite{badenes2007,yamaguchi2015}, searching for surviving companion stars in supernova remnants \cite{kerzendorf2009,kerzendorf2012,schaefer2012}, galactic chemical evolution \cite{seitenzahl2013b,kobayashi2015} or radio emission from potential interaction with the circumstellar medium \cite{horesh2012,chomiuk2012,Torres2014a}. However, the question of the nature of the progenitor systems of spectroscopically ``normal" SNe Ia remains unanswered.  
A Galactic SN Ia would obviously be of immense value in settling the debate, at least for that particular event.  
The Galactic SN Ia rate as given by Adams \emph{et al.} \cite{2013ApJ...778..164A} of $1.4^{+1.4}_{-0.8}$ per century is 30\% of the total supernovae rate, and the same authors give the most probable distance to a Galactic SN Ia as $d = 9\;{\rm kpc}$.  We will use 10 kpc as a standard distance.

An observational signal that could help bring clarity to the SN Ia progenitor and explosion mechanism debate is the neutrino signal produced by a SN Ia \cite{Odrzywolek2011a}. Neutrinos from a core-collapse supernova were observed in  1987 \cite{Alekseev:1987JETPL..45..589A,Bionta:1987qt,Hirata:1987hu} and, despite its paucity, the signal was fully exploited in order to extract competitive limits on multiple neutrino properties as well as testing the basic paradigm of core-collapse. 
Should the next burst from a core-collapse supernova arrive tomorrow, many more events will be recorded for the very simple reason that, compared to the size and scale of detectors operating in 1987, present-day detectors such as Baksan \cite{Baksan2012}, Super-Kamiokande \cite{Ikeda:2007sa}, LVD \cite{2007APh....27..254A}, KamLAND \cite{2000NuPhS..87..312D}, MiniBOONE \cite{2002PhRvD..66a3012S}, Borexino \cite{Cadonati:2000kq}, Daya Bay \cite{2016DayaBay} and the dedicated supernova burst detector HALO \cite{2008JPhCS.136d2077D}  are much larger and/or more sensitive to lower energies or to a broader set of channels.
The burst would also be recorded in IceCube \cite{2009arXiv0908.0441K} and Antares \cite{2008arXiv0810.1416D}  but with no event-by-event energy resolution.  Future detectors such as Hyper-Kamiokande \cite{Abe:2011ts}, DUNE \cite{Acciarri:2015uup}, JUNO \cite{An:2015jdp}, and KM3NeT \cite{2009KM3NeT} will have larger statistics and even broader flavor sensitivity. In comparison to core-collapse supernovae, the flux of neutrinos from SNe Ia is smaller by about four orders of magnitude and the spectrum has a lower mean energy. On the plus side, the relatively low flux makes computing the flavor transformation through the supernova simple: the only effect one needs to include is the effect of matter. The neutrino self-interaction effect \cite{Duan:2006an,Duan:2006jv} does not occur. However, as in core-collapse supernovae, the matter effect is not stationary over the duration of the neutrino burst and models of SNe Ia show the star does not explode with spherical symmetry and so one might expect some degree of line-of-sight dependence. 

The goal of this paper is to compute the signal from a deflagration-to-detonation transition SN Ia as completely as possible by including the time, energy, and line-of-sight dependence of the flavor evolution through the supernova and the time dependence of the neutrino spectrum. The simulation we adopt is the DDT SN Ia by Seitenzahl \emph{et al.} \cite{Seitenzahl2015a}. We restrict our attention to this one particular model in order to explain the many details of the calculation and leave alternative explosion mechanisms for future investigation. The paper will proceed as follows: in Section \S\ref{sec:SneModel} we describe the particular DDT SN Ia model used, while Section \S\ref{sec:Production}  describes how the neutrino spectrum is computed. In Section \S\ref{sec:NeutrinoOscillation} we show how neutrino oscillations are taken into account and briefly describe the various oscillation phenomena that can occur. The detection of the neutrinos on Earth-based detectors is discussed in Section \S\ref{sec:NeutrinoDetection} and we conclude with Section \S\ref{sec:Conclusion}.


\section{Supernova Simulation \label{sec:SneModel}}
The first step in computing the neutrino signal from the DDT scenario for SNe Ia is a 
simulation. The particular simulation explored here is the N100$\nu$ model described in detail in \cite{Seitenzahl2012a} and \cite{Seitenzahl2015a}. We include a short description for completeness. The key feature of this model is that the deflagration-to-detonation transition is delayed. This delay allows the deflagration flame to produce enough intermediate mass elements before the detonation takes over \cite{Khokhlov1989}. The initial stellar setup has a central density of $\rho\approx3\times10^9 \text{ g/cm}^3$, a mass of $M=1.4\text{ M}_\text{Sun}$, a radius of $R\approx2\times10^8\text{ cm}$ and is setup as a cold  $\left(T=5\times10^5\text{ K}\right)$ carbon oxygen white dwarf. This stellar setup is then hydrodynamically evolved using the thermonuclear supernova code \textsc{Leafs}. The initial deflagration is seeded and the transition to detonation is modeled stochastically \cite{Ciaraldi-Schoolmann2013}. From the neutrino perspective, the densities of the stellar material are not high enough to trap them. Thus the WD is transparent to neutrinos and the N100$\nu$ model takes this internal energy loss due to neutrino emission into account dynamically.

\begin{figure}[ht]
	\centering
    \begin{minipage}{0.5\textwidth}
        \centering
        \includegraphics[width=1\linewidth]{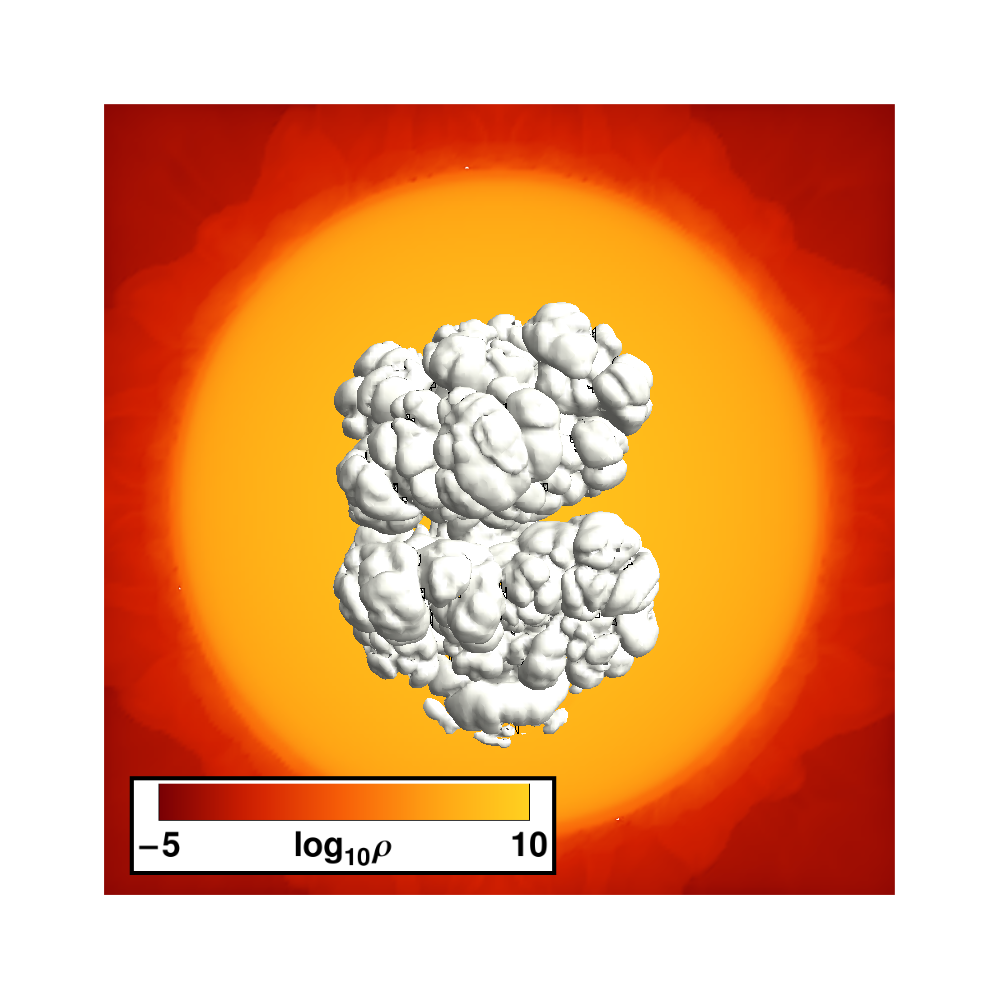}
    \end{minipage}
        \caption{Density plot of N100$\nu$ model at $t=0.8$ s. The white surface shows the location of the deflagration flame front, separating the nuclear ashes from the fuel.}
        \label{fig:DensityN100v0.8s}
\end{figure}

Figure (\ref{fig:DensityN100v0.8s}) shows the density of the SN at $t=0.8s$. The white contours represent the edges of the 3D deflagration flame surface and the colored areas represent the 2D density of the star for the $z=0$ slice. The deflagration contours represent discontinuities in the density, which need to be accounted for to accurately model neutrino propagation through the star. It is also in the hot zones consisting of deflagration ash where the majority of the neutrinos are produced.


\section{Neutrino Production \label{sec:Production}}
While the N100$\nu$ model does compute the energy loss to neutrinos, it does not compute the neutrino emission spectrum. In order to compute the neutrino spectrum we post-process the simulation using the software package \textsc{NuLib} \cite{Sullivan2015}. \textsc{NuLib} is an open-source neutrino interaction library that can be used to calculate neutrino emissivities, scattering opacities, and absorption opacities. The stellar equation of state (EOS) used in our implementation of \textsc{NuLib} is calculated in \cite{Steiner2013} and translated for \textsc{NuLib} by \cite{OConnor2010}. Weak interactions are calculated in \textsc{NuLib} via rates tables from \cite{Fuller1982,Oda1994,Langanke2000,Langanke2003} and an approximation scheme for the spectrum is described in \cite{Sullivan2015}. Thermal neutrino pair production spectra are calculated using the equations derived in \cite{Burrows2006a}. 

\subsection{Calculation Strategy}
The N100$\nu$ model gives the following data at each point on a $512\times512\times512$ Cartesian grid: nuclear pseudo-abundances, density, temperature, and electron fraction. This set of data is used to set up the EOS which is then used to calculate the weak and thermal neutrino emissivities. While the SFHo \cite{Steiner2013} EOS is designed to describe core-collapse supernova environments, it is valid for any region that is in nuclear statistical equilibrium (NSE) where $T>3\times10^9\text{K}$. The material in NSE \cite{Seitenzahl2009} accounts for the vast majority of the neutrino emission. Thus our calculation strategy is to only compute the neutrino emissivity from NSE zones from the N100$\nu$ model and ignore the remainder. We shall show that the `NSE only' strategy gives neutrino luminosities in good agreement with those calculated by the N100$\nu$ model itself.

\subsection{Neutrino Processes\label{subsec:NeutrinoProcesses}}
The many processes that could produce neutrinos are usually divided into weak and thermal processes. Weak processes only produce electron and anti-electron flavor neutrinos and thermal processes produce neutrinos of all flavors. The weak processes that are included here are
\begin{align}
	p+e^-&\rightarrow n+\nu_e,\label{eqn:weak1}\\
    n+e^+&\rightarrow p+\overline{\nu}_e,\label{eqn:weak2}\\
    e^-+\left(A,Z\right)&\rightarrow\left(A,Z-1\right)+\nu_e,\label{eqn:weak3}\\ 
    e^++\left(A,Z-1\right)&\rightarrow\left(A,Z\right)+\overline{\nu}_e,\label{eqn:weak4}
\end{align}
$\beta^\pm$ decays are not considered because the time window of significant neutrino emission is shorter than the decay time. 
Furthermore $\beta^-$ decays are often Fermi-blocked because of electron degeneracy. 
The weak rates are heavily dependent on the composition of the material, which is itself heavily dependent on the density, temperature, and electron fraction.

There are many different sources of thermal neutrinos in stellar environments, pair, photo, plasma, Bremsstrahlung, and recombination. Each dominates under different circumstances but for NSE material, pair neutrinos arising from electron positron annihilation are the dominant contributor
\begin{align}
	e^-+e^+\rightarrow\nu_{e,\mu,\tau}+\overline{\nu}_{e,\mu,\tau}.\label{eqn:thermal}
\end{align}
Only pair thermal neutrinos are included in our calculations. Thermal neutrinos are especially important because only thermal processes can produce neutrinos of $\mu$ and $\tau$ flavor. However, weak processes greatly dominate over thermal processes during the periods of maximum neutrino emission.

\begin{figure}[ht]
	\centering
    \begin{minipage}{0.99\textwidth}
        \centering
        \includegraphics[trim={0 0 0 3.2cm},clip,width=1\linewidth]{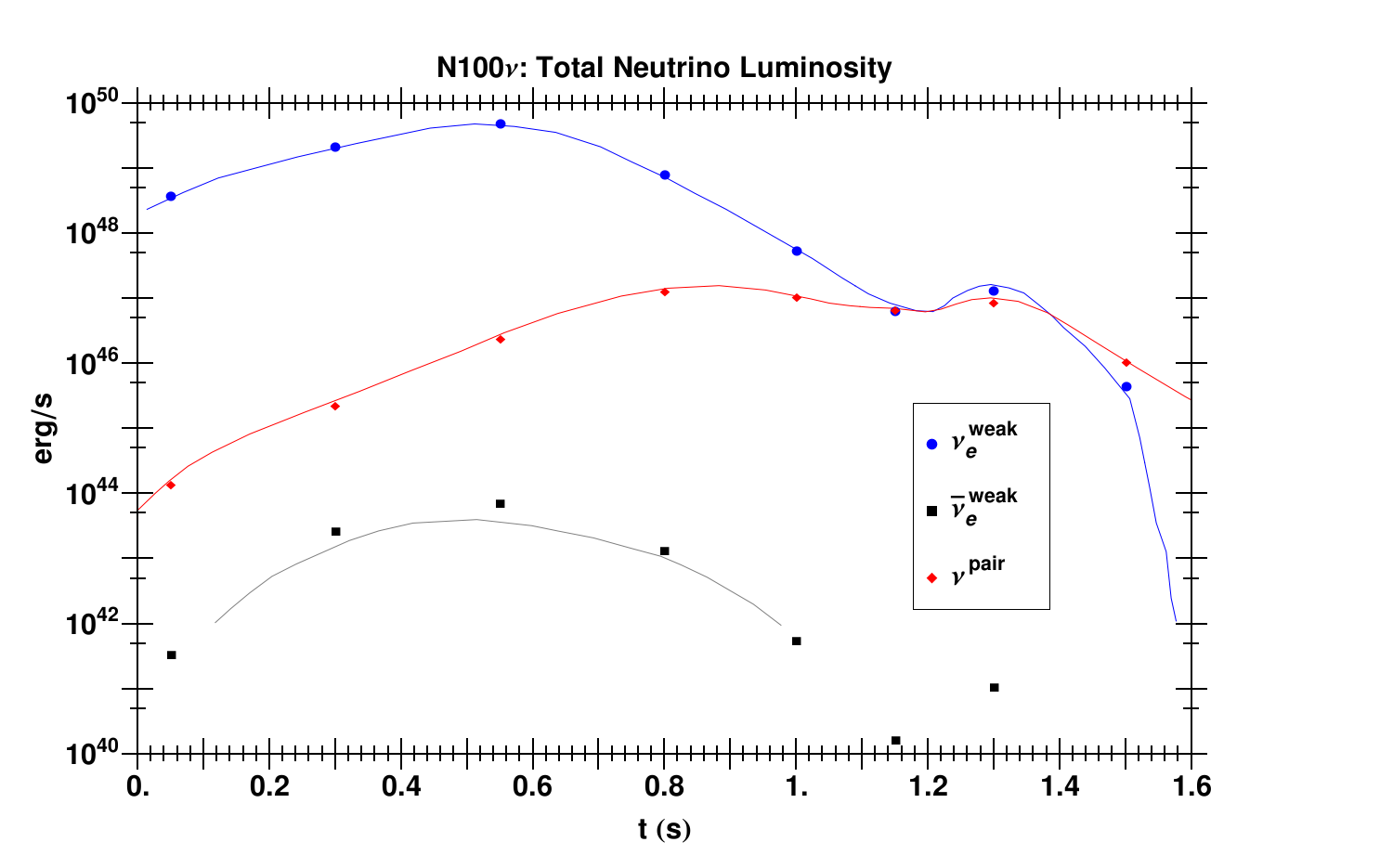}
    \end{minipage}
        \caption{Total neutrino luminosity from N100$\nu$ model as a function of elapsed time. The different colors represent the different nuclear processes that contribute to the luminosity. The points are calculated via \textsc{NuLib} and the lines are those calculated in \cite{Seitenzahl2015a}.}
        \label{fig:MeVsIvo}
\end{figure}

As a validation of the `NSE only' strategy with only the nuclear processes described by Equations (\ref{eqn:weak1} - \ref{eqn:thermal}), consider Figure (\ref{fig:MeVsIvo}).  Figure (\ref{fig:MeVsIvo}) plots three total luminosities for the energy range $0.01<E_\nu<100\text{ MeV}$. The blue dots represent all $\nu_e$ luminosity arising from processes (\ref{eqn:weak1} \& \ref{eqn:weak3}). The black squares represent all $\overline{\nu}_e$ luminosity arising from processes (\ref{eqn:weak2} \& \ref{eqn:weak4}). The red diamonds represent all $\nu$ and $\overline{\nu}$ luminosity arising from process (\ref{eqn:thermal}). The blue, black and red lines are those computed in \cite{Seitenzahl2015a}. The agreement between the values calculated via \textsc{NuLib}, and the values dynamically calculated in the N100$\nu$ model, is remarkable. 
The agreement is less good for $\overline{\nu}_e$, which is a reflection of the fact that \cite{Seitenzahl2015a} included $\beta^\pm$ decays in the NSE material by implementing the tables of \cite{Seitenzahl2009}. The small differences are not worrisome because $\overline{\nu}_e$ emission is dominated by pair production, which has much better agreement. Not including $\beta^\pm$ decays is therefore justified. One important feature is that the \textsc{NuLib} rates also reveal the double peak structure present in the N100$\nu$ model \cite{Seitenzahl2015a}. The second peak represents the DDT and its delay until $1.3\,\mathrm{s}$ is a crucial feature of the N100$\nu$ model.

\subsection{Neutrino Production Spectral Results}
\begin{figure}[ht]
	\centering
    \begin{minipage}{0.99\textwidth}
        \centering
        \includegraphics[trim={0 0 0 1.2cm},clip,width=1\linewidth]{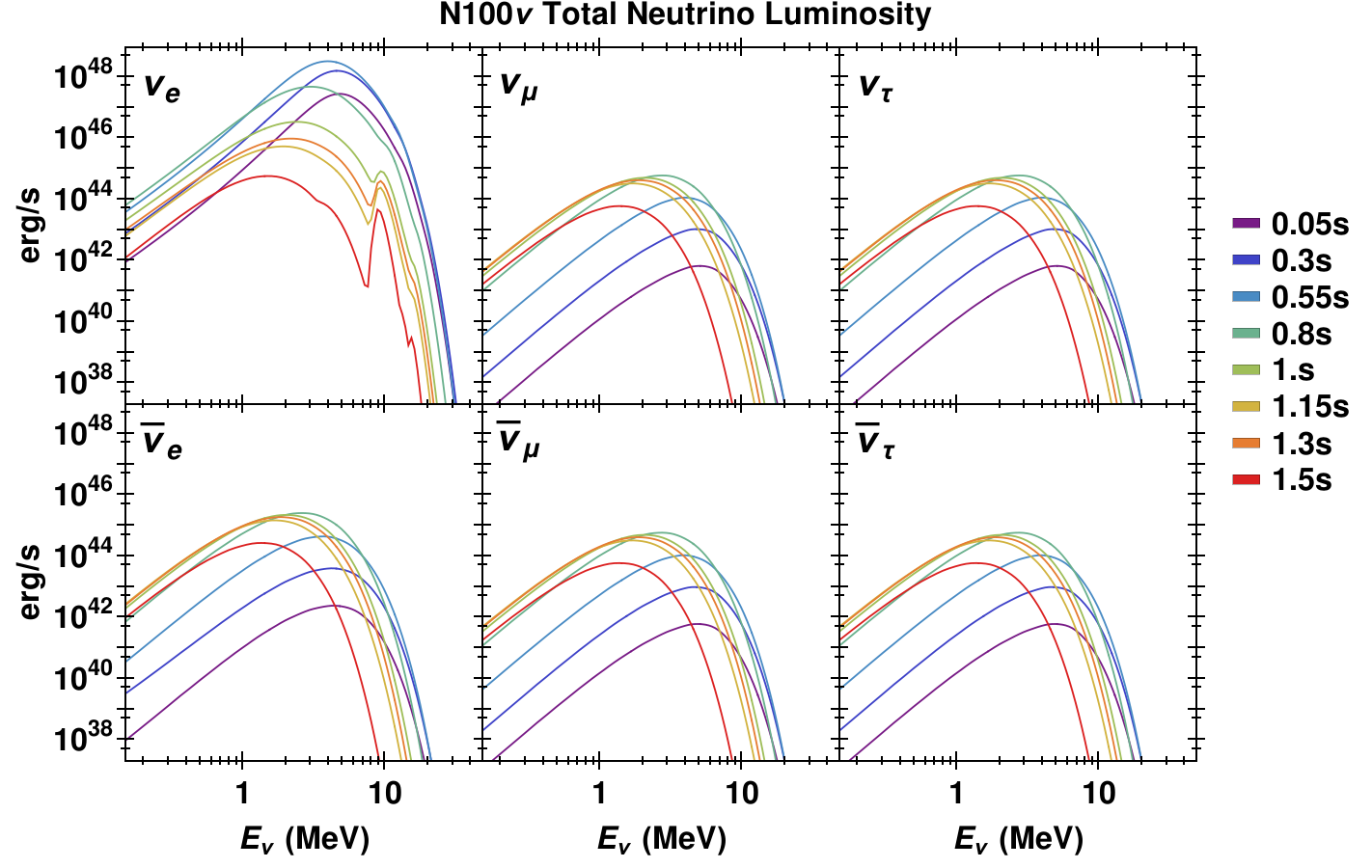}
    \end{minipage}
        \caption{Total neutrino luminosity from the N100$\nu$ model. The curve represents the sum of the NSE contributions from the processes described by Equations (\ref{eqn:weak1} - \ref{eqn:thermal}).}
        \label{fig:N100vTotalNeutrinoLuminosity}
\end{figure}

The results of the neutrino luminosity spectral calculation are shown in Figure (\ref{fig:N100vTotalNeutrinoLuminosity}) calculated by post-processing the N100$\nu$ data through \textsc{NuLib}. Only the processes described by Equations (\ref{eqn:weak1} - \ref{eqn:thermal}) are included and only the contributions from NSE ($T>3\times 10^9\text{ K}$) zones are summed.

Figure (\ref{fig:N100vTotalNeutrinoLuminosity}) reveals similarities in the luminosities of $\nu_\mu$, $\overline{\nu}_\mu$, $\nu_\tau$ and $\overline{\nu}_\tau$. These similarities are because the emissivities of these flavors are completely dominated by thermal emission. Thermal emission mostly dominates the $\overline{\nu}_e$ emission as well, but the contributions from the processes described in Equations (\ref{eqn:weak2}) \& (\ref{eqn:weak4}) are significant, especially at low energy ($<1\text{ MeV}$) and early times ($<0.5 \text{ s}$). Lastly, the $\nu_e$ luminosity is is dominated by weak processes at early times ($<1 \text{ s}$) and after that, weak and thermal processes are equally important.

The most interesting feature of Figure (\ref{fig:N100vTotalNeutrinoLuminosity}) is the $\nu_e$ luminosity spectrum. Firstly, at all times and energies the $\nu_e$ luminosity is orders of magnitude greater than all the other flavors. This is not surprising because the explosive nucleosynthesis in SNe Ia populates the proton-rich side of the valley of stability in the nuclear chart,
and thus the processes described by Equations (\ref{eqn:weak1}) \& (\ref{eqn:weak3}) dominate. 

\begin{figure}[ht]
	\centering
    \begin{minipage}{0.99\textwidth}
        \centering
        \includegraphics[trim={0 0 0 1.5cm},clip,width=1\linewidth]{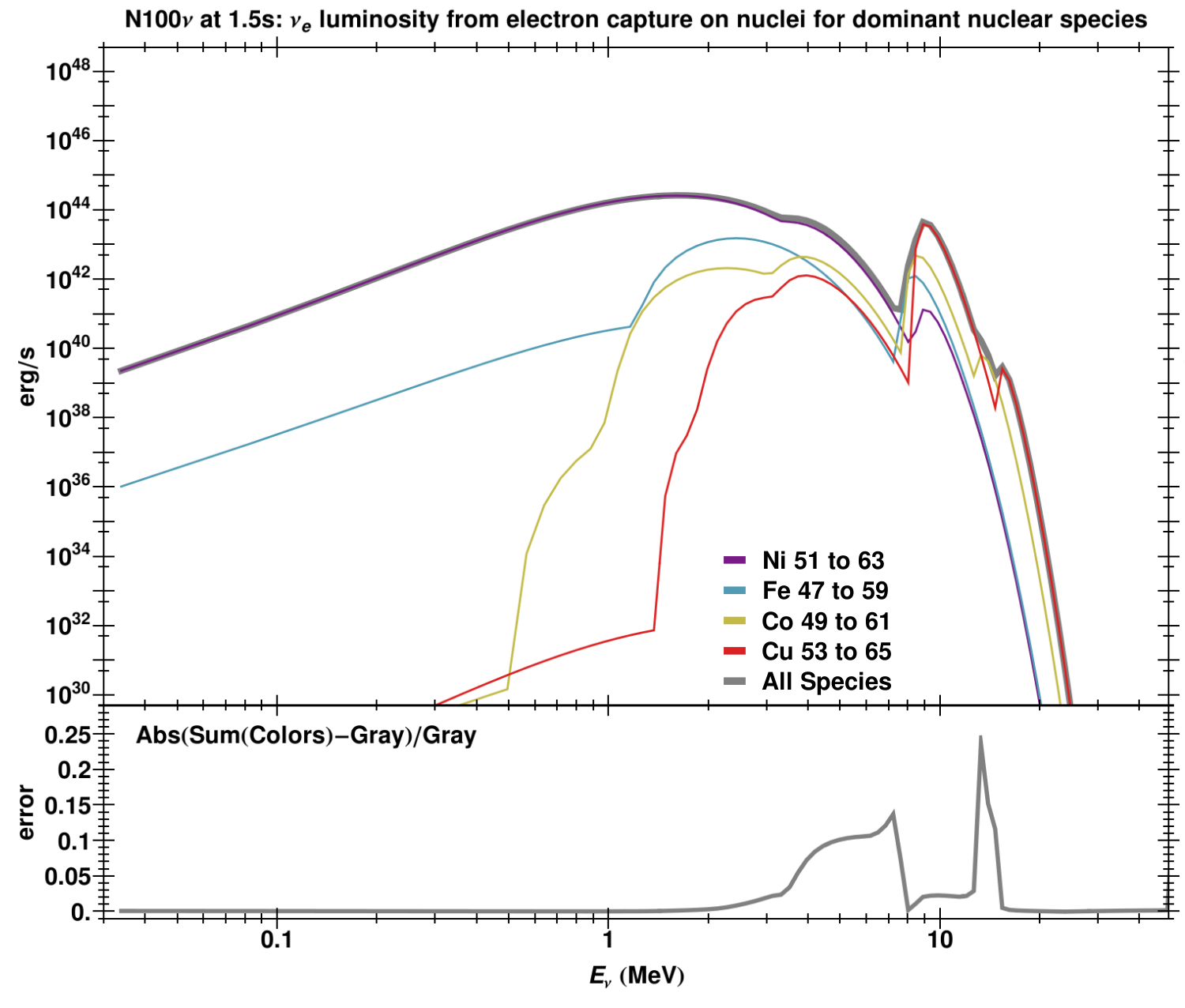}
    \end{minipage}
        \caption{$\nu_e$ luminosity from electron capture on nuclei for dominant nuclear species for the $t=1.5\text{ s}$ time slice. The colored curves represents specific species contributions and the gray curve represents the total contribution from all 8140 species considered by \textsc{NuLib}. The bottom graph plots the error.}
        \label{fig:N100vLuminosityPerSpecies}
\end{figure}

The next feature of interest in the $\nu_e$ luminosity spectrum is the 10 MeV peak that begins to form at $t\approx 1\;\text{s}$ into the explosion. This peak is the most notable feature and its source is the weak process described in Equation (\ref{eqn:weak3}). Figure (\ref{fig:N100vLuminosityPerSpecies}) shows which nuclei are responsible for the spectral shape of the $\nu_e$ luminosity. Below $\sim7\text{ MeV}$ the luminosity is dominated by nickel ($^{56}\text{Ni}$ below $\sim3\text{ MeV}$ and $^{55}\text{Ni}$ between $\sim3\text{ MeV}$ and $\sim7\text{ MeV}$). But above $\sim7\text{ MeV}$ the contributions from iron, cobalt, and then copper dominate. The important point is that the 10 MeV peak is mostly caused by electron capture on copper. In particular, the isotopes $^{57}\text{Cu}$, $^{53}\text{Co}$, and $^{51}\text{Fe}$ are responsible for the 10 MeV peak. 

The 10 MeV peak is a very interesting feature because if it is ever experimentally seen it could give information about the nuclear composition of the SN in a neutrino signal. 
In Section \S\ref{sec:NeutrinoDetection} we shall pay particular attention to the question of whether this feature can be observed. 
However, before these neutrinos can be detected on Earth, they must first oscillate through the SN material and traverse the interstellar medium, which for our purposes can be approximated as vacuum.


\section{Neutrino Oscillation \label{sec:NeutrinoOscillation}}
The flavor structure of the neutrino flux through a detector on Earth will not be the flavor composition of the flux at the source. This phenomenon of neutrino-flavor oscillations will modify the flavor content and one must take the change into account before one can predict the event rate in a given detector. The oscillations depend on many factors, in particular, the density and electron fraction of the medium through which the neutrinos propagate. These quantities evolve with time as the supernova proceeds leading to new time dependent features in the flux. The second process that needs to be taken into account is the decoherence of the neutrino through the vacuum. Decoherence occurs because the propagation distance is much larger than the coherence length \cite{Giunti1998}. The one flavor transformation process that we do not need to include is neutrino self interactions - see \cite{2009JPhG...36k3201D,2010ARNPS..60..569D} for reviews of this very interesting physics. The reason is that the neutrino flux is too small for self-interactions to play a significant role. 

\subsection{Theoretical setup\label{subsec:Theory}}
The quantum mechanical phenomenon of neutrino oscillations is a consequence of the mismatch between the neutrino mass eigenstates and interaction/flavor eigenstates. The evolution matrix $S^{(YX)}(r_2,r_1)$ relates the neutrino states in some basis $(X)$ at some initial position $r_1$ to the states in a possibly different basis $(Y)$ at $r_2$. The transition probabilities from some state $x$ in $(X)$ to state $y$ in $(Y)$, denoted by $P^{(YX)}_{yx}(r_2,r_1)$, are calculated from the elements of $S^{(YX)}$ by $P^{(YX)}_{yx} = |S^{(YX)}_{yx}|^2$. 
The two bases we refer to are the flavor and mass bases. 
The two bases are related by a unitary "mixing" matrix $U$ which is parametrized in terms of 
three mixing angles, $\theta_{12}$, $\theta_{23}$, $\theta_{13}$, and a CP-violating phase $\delta_{CP}$. Other possible phases in the standard paradigm do not influence the outcome of neutrino oscillations. In terms of these parameters $U$ is 
	\begin{align*}
		U=
		\begin{pmatrix}
		1 & 0 & 0 \\
		0 & c_{23} & s_{23} \\
		0 & -s_{23} & c_{23}
		\end{pmatrix} 
		\begin{pmatrix}
		c_{13} & 0 & s_{13}e^{-i\delta_{CP}} \\
		0 & 1 & 0 \\
		-s_{13}e^{i\delta_{CP}} & 0 & c_{13} 
		\end{pmatrix}
		\begin{pmatrix}
		c_{12} & s_{12} & 0 \\
		-s_{12} & c_{12} & 0 \\
		0 & 0 & 1
		\end{pmatrix},
	\end{align*}
with $c_{ij}=\cos\theta_{ij}$ and $s_{ij}=\sin\theta_{ij}$. 
The evolution matrix in any basis can be computed from the Schrödinger equation

\begin{equation}
\imath \frac{dS^{(XX)}}{dr} = H^{(X)} S^{(XX)} \label{eqn:Schroedinger}
\end{equation}
The Hamiltonian $H$ is the sum of two terms: a vacuum term $H_V$ and a matter term $H_M$. 
The vacuum Hamiltonian in the flavor basis depends upon the neutrino energy $E$ and is given by
\begin{equation}
H^{(f)}_V = \frac{1}{2E}\,U \begin{pmatrix}
		m_1^2 & 0 & 0 \\
		0 & m_2^2 & 0 \\
		0 & 0 & m_3^2
		\end{pmatrix} U^{\dagger}
\end{equation}
with $m_i$ the neutrino masses. 
The oscillation parameter values chosen for these and all results in this paper are 
\begin{align}
\left( m^2_2-m^2_1,|m^2_{3}-m^2_{2}|,\theta_{12},\theta_{13},\theta_{23},\delta_{cp}\right)=\left(7.5\times10^{-5}\text{eV}^2,2.32\times10^{-3}\text{eV}^2,33.9^\circ,9^\circ,45^\circ,0\right).
\end{align}
We shall explore both signs for the difference $m^2_{3}-m^2_{2}$: the positive difference is the normal mass ordering (NMO) and the negative choice the inverted mass ordering (IMO). 

The matter Hamiltonian arises due to a difference between the interaction of electron flavor neutrinos/antineutrinos with the medium compared to the $\mu$ and $\tau$ flavors.  
The interaction can be described by an effective potential \cite{1978PhRvD..17.2369W,Mikheyev:1985aa} which leads to the Hamiltonian in the flavor basis given by 
\begin{equation}
H^{(f)}_M = \sqrt{2}\,G_F\,n_e \begin{pmatrix}
		1 & 0 & 0 \\
		0 & 0 & 0 \\
		0 & 0 & 0
		\end{pmatrix} 
\end{equation}
where $G_F$ is the Fermi constant and $n_e$ is the electron number density.
The electron density can be rewritten as $n_e = Y_e n_N$ where $Y_e$ is the electron fraction and $n_N$ the nucleon density. The electron fraction and the nucleon density are provided by the simulation. 
But before inserting $Y_e(r)$ and $n_N(r)$ into a neutrino evolution code, we must take care 
to correctly insert the discontinuities due to both the deflagration and detonation flame fronts. It has been shown by Lund \& Kneller \cite{Lund2013} that a failure to properly account for discontinuities leads to errors in the transition probabilities. 

The neutrino evolution is computed using the code \textsc{Sa}. While it is possible to solve the neutrino flavor evolution using Equation (\ref{eqn:Schroedinger}) in any basis, in practice the neutrinos propagate over such large distances compared to the oscillation length that it can become very computationally expensive if the basis is not chosen wisely. Efficiency can be greatly improved by moving to the adiabatic basis as described in \cite{Kneller2009a}. Working in this basis, the evolution matrix $S$ is parametrized by a set of eleven variables; three adiabatic phases and eight variables to describe the unitary matrix that accounts for the departure from the adiabatic solution. We solve the set of differential equations from the center of the simulation along a given ray through the simulation to the edge of the data. The transition probabilities between matter basis states are the most suitable for describing the neutrino evolution through the supernova because a) the transition probabilities do not depend upon the exact point where one stops the calculation (as would occur in the flavor basis) and b) the matter states are the local eigenstates of the neutrino, so one can describe the evolution as being adiabatic or diabatic depending on whether the survival probabilities, $P^\text{(m)}_{\text{ii}}$, are close to unity or zero respectively. The evolution matrix (and the associated transition probabilities) in any other basis can be obtained by applying suitable unitary transformations at either the initial or final point of the integration. As a reference, the matter and flavor basis states closely align in dense matter. In the normal mass ordering we find an approximate equivalence between: $\nu_1 \approx \nu_{x}$, $\nu_2 \approx \nu_{y}$ and $\nu_3 \approx \nu_{e}$, $\bar{\nu}_1 \approx \bar{\nu}_{e}$, $\bar{\nu}_2 \approx \bar{\nu}_{x}$ and $\bar{\nu}_3 \approx\bar{\nu}_{y}$ where $x$ and $y$ denote a mixture of $\nu_{\mu}$ and $\nu_{\tau}$. In the inverted mass ordering the approximate equivalence is between: $\nu_1 \approx \nu_{x}$, $\nu_2 \approx \nu_{e}$ and $\nu_3 \approx \nu_{y}$, $\bar{\nu}_1 \approx \bar{\nu}_{x}$, $\bar{\nu}_2 \approx \bar{\nu}_{y}$ and $\bar{\nu}_3 \approx\bar{\nu}_{e}$.

After the neutrinos emerge from the supernova, the wavefunction decoheres as the neutrino travels to Earth. 
The decoherence means the probability that some initial neutrino flavor $\beta$ produced in the center of the supernova is detected on Earth as flavor $\alpha$ is given by
\begin{align}
    P_{\alpha\beta}=P_{\nu_\alpha\rightarrow\nu_\beta}=\sum_i |U_{\beta i} |^2 P^\text{(mf)}_{i\alpha}\left(R_*,R_\nu\right),\label{eqn:Decoherence}
\end{align}
where $R_*$ represents the radius of the outer edge of the supernova, $R_\nu$ represents the radius of the neutrino production point (near the center of the supernova which we take to be zero), and $P^\text{(mf)	}_{i\alpha}\left(R_*,R_\nu\right)$ is the probability that a neutrino in some initial flavor state $\alpha$ would have been detected as mass state $i$ as it traveled from $R_\nu$ to $R_*$ (assuming $R_*$ is the vacuum). 


\subsection{Numerical Oscillation Results\label{subsec:OcsResults}}

\begin{figure}[ht]
	\centering
    \begin{minipage}{0.49\textwidth}
        \centering
        \includegraphics[trim={0 0 0 2.2cm},clip,width=1\linewidth]{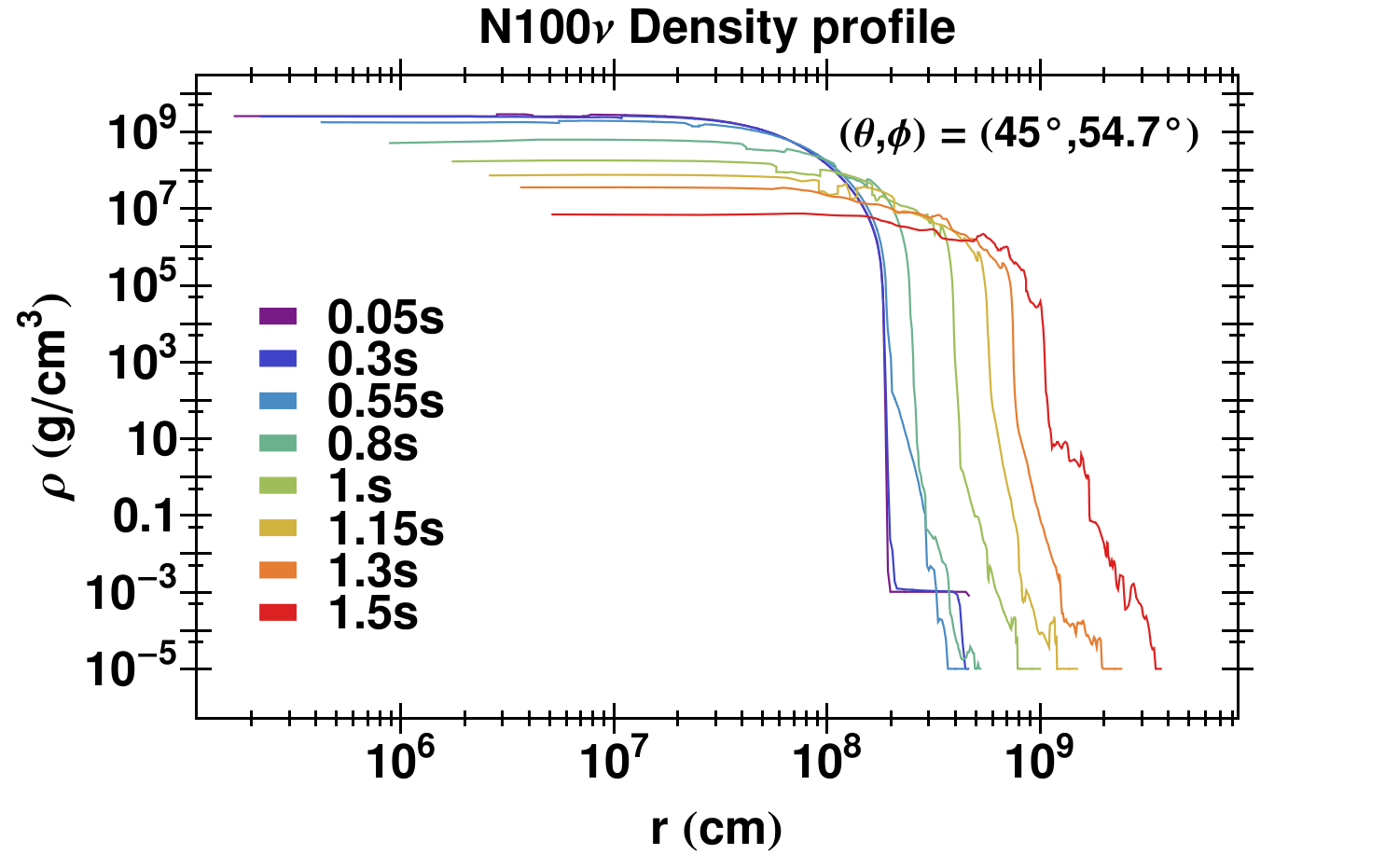}
    \end{minipage}
    \begin{minipage}{0.49\textwidth}
        \centering
        \includegraphics[trim={0 0 0 2.2cm},clip,width=1\linewidth]{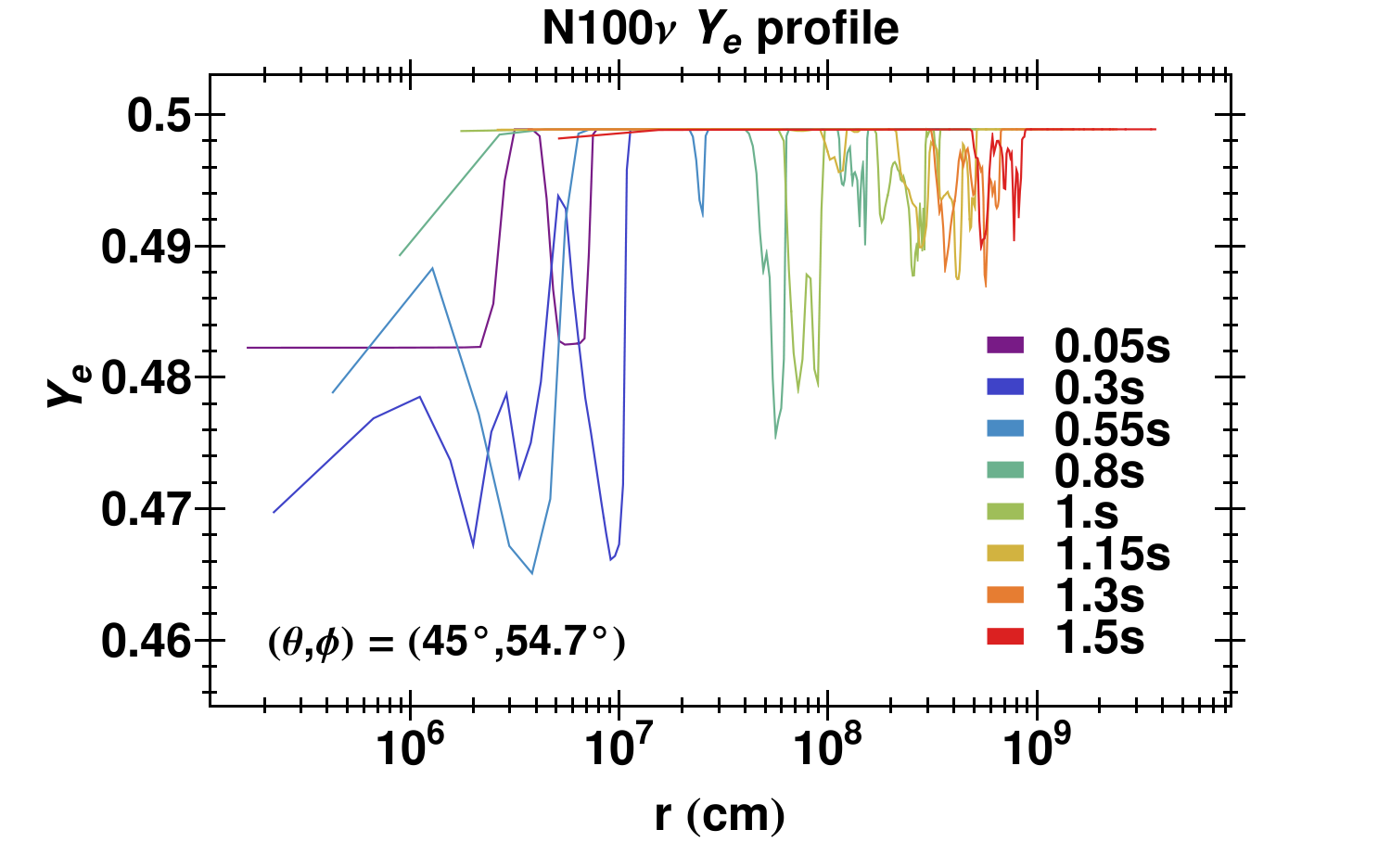}
    \end{minipage}
        \caption{Density and $Y_e$ profiles of the N100$\nu$ model for a particular choice of zenith and azimuth angles.}
        \label{fig:RhoYeProfile}
\end{figure}
\begin{figure}[t!]
\includegraphics[trim={0 0 0 1.9cm},clip,width=0.93\linewidth]{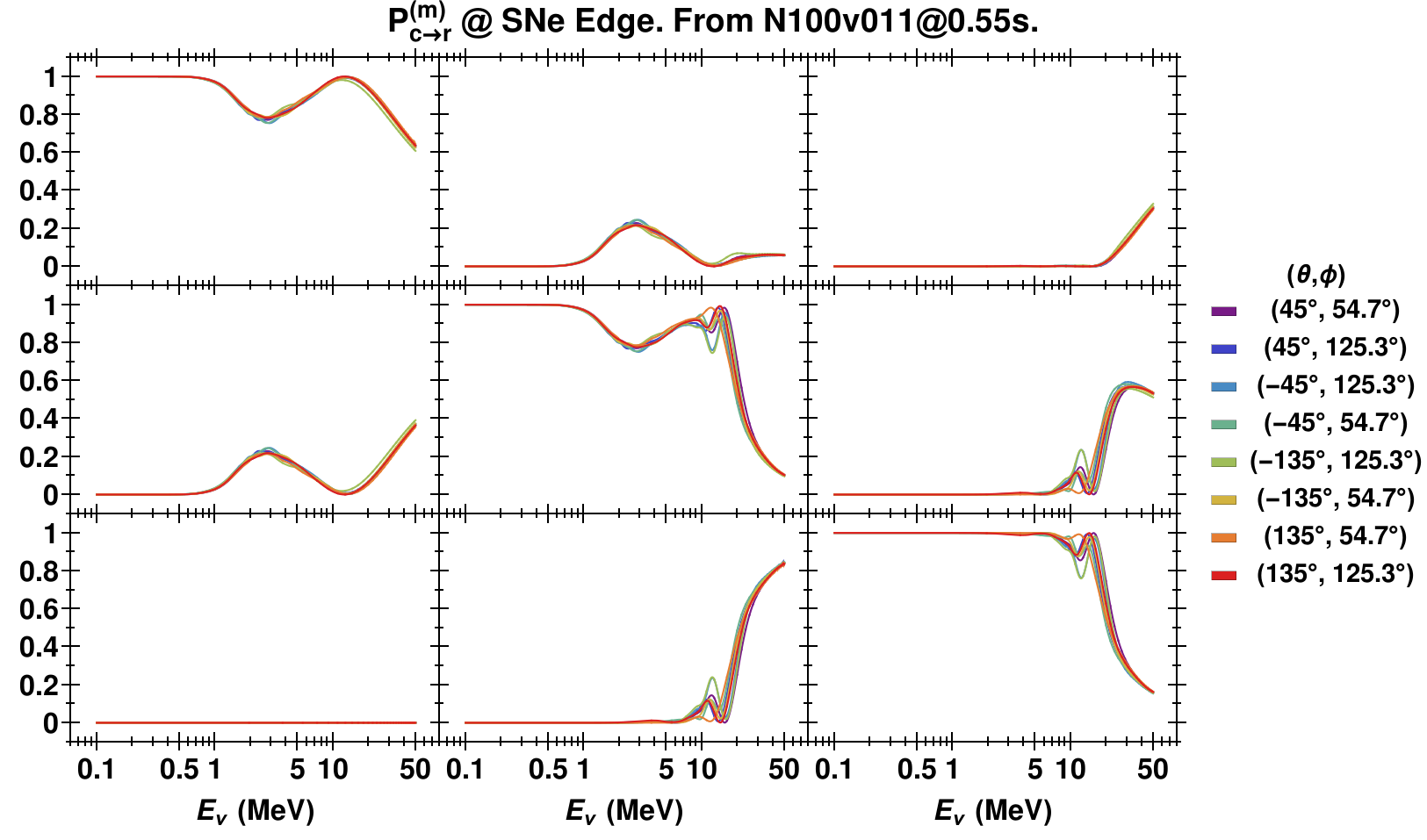}
\caption{The matter basis transition probabilities $P^\text{(m)}_{\text{cr}}\left(E_\nu\right)$ where r and c denote a row and column position in the 3x3 plot grid. The mass ordering is normal and the time-slice is for the $t=0.55\;\text{s}$ time-slice. The different colored lines correspond to the different zenith and azimuth angles as given in the legend.}
\label{fig:N100vSmNHAt0.55s}
\end{figure}
\begin{figure}[b!]
\includegraphics[trim={0 0 0 1.9cm},clip,width=0.93\linewidth]{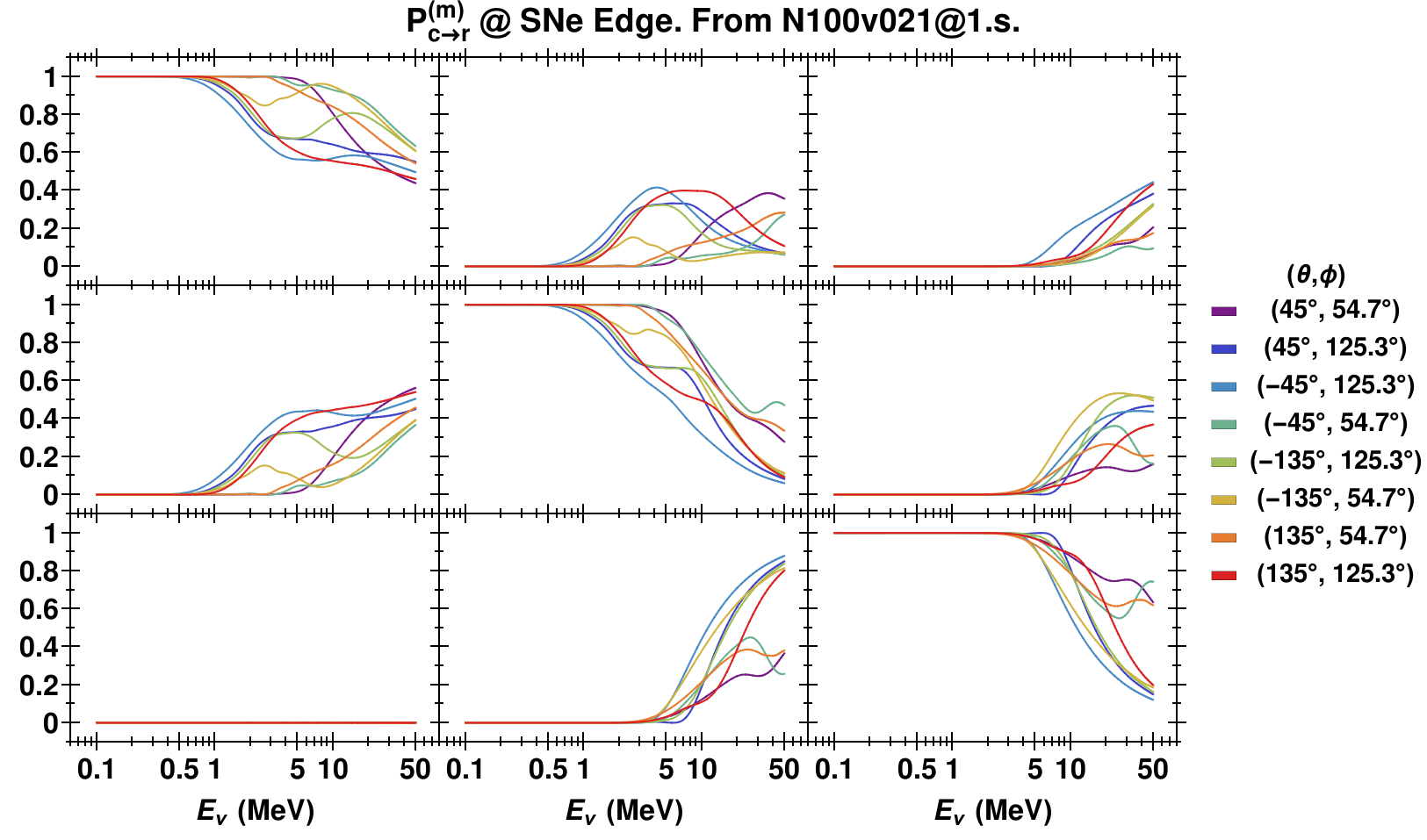}
\caption{The same as Figure (\ref{fig:N100vSmNHAt0.55s}) but for the $t=1\;\text{s}$ time-slice.}
\label{fig:N100vSmNHAt1s}
\end{figure}
\begin{figure}[b!]
\includegraphics[trim={0 0 0 1.9cm},clip,width=1\linewidth]{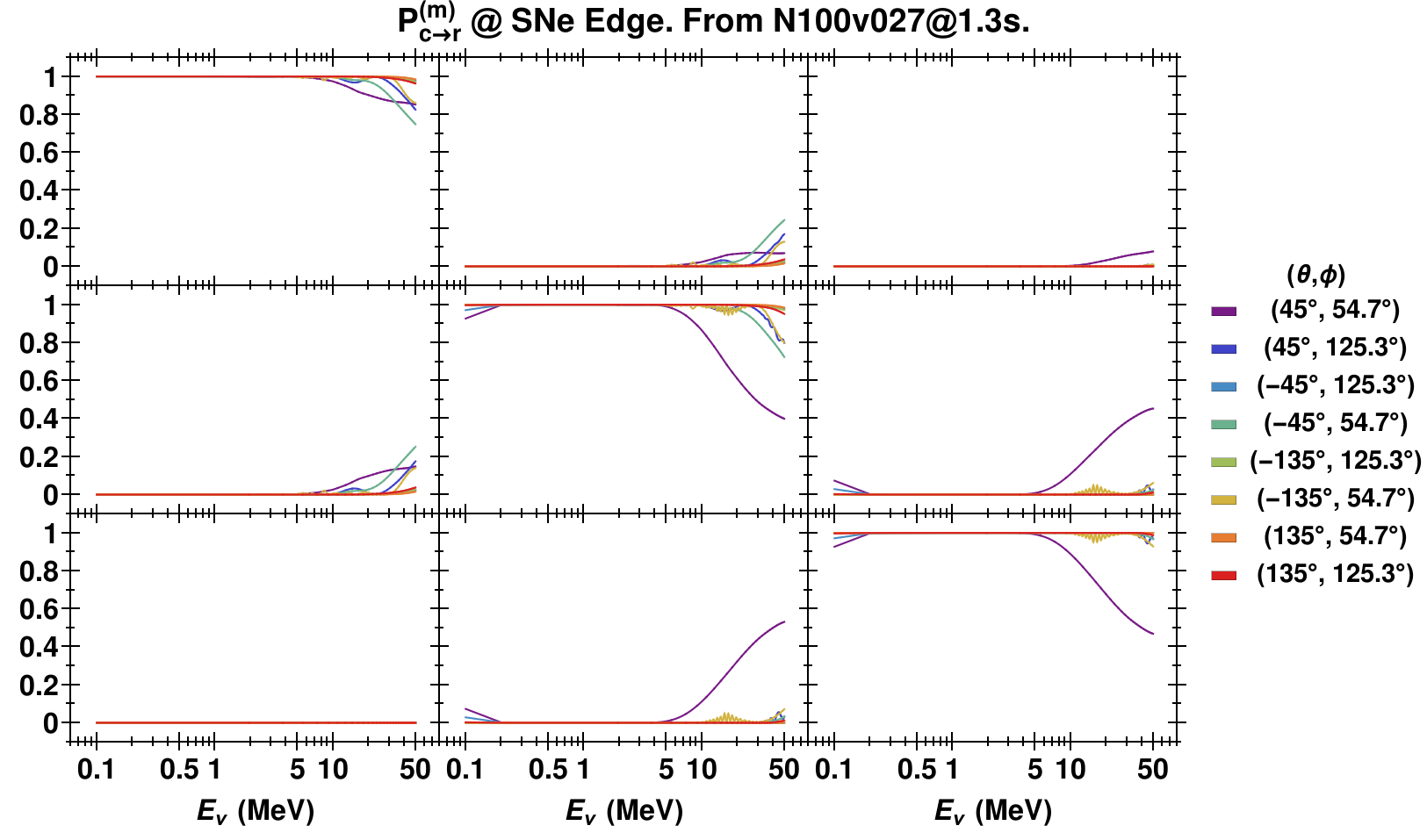}
\caption{The same as Figure (\ref{fig:N100vSmNHAt0.55s}) but for the $t=1.3\;\text{s}$ time-slice.}
\label{fig:N100vSmNHAt1.3s}
\end{figure}

The flavor transformation from the center of the supernova to the vacuum will depend upon the matter density and the electron fraction along the neutrino trajectory. As we have seen, the supernova explosion is not spherically symmetric which means that we might expect a dependence upon the specific line of sight chosen. In order to explore whether there is significant line-of-sight dependence we select eight different rays through the simulation corresponding to the trajectories which start at the center of the SN and propagating along the `diagonals' i.e. along the corners of a cube centered on the SN. Figure (\ref{fig:RhoYeProfile}) displays the density and electron fraction along one of these 8 trajectories, denoted by $\left(\theta,\phi\right)=\left(45^\circ,54.7^\circ\right)$. The different colors show how the profiles change as the SN evolves. The most important feature to note is the steep drop in density at the edge of the star at early epochs. As the star explodes the density gradient at the edge of the star softens considerably and fluctuations start to appear. 

The matter basis transition probabilities $P^\text{(m)}_{\text{ij}}$ as a function of energy at three snapshot times are shown in Figures (\ref{fig:N100vSmNHAt0.55s}) through (\ref{fig:N100vSmNHAt1.3s}) for the case of the normal mass ordering. The snapshot times are chosen to be $t=0.55\;{\rm s}$ corresponding to the peak of the neutrino luminosity, $t=1.0\;{\rm s}$ during the period with the greatest line-of-sight dependence, and $t=1.3\;{\rm s}$ corresponding to the secondary peak in luminosity. 
A number of features can be seen in the figures:
\begin{itemize}
\item At early times, Figure (\ref{fig:N100vSmNHAt0.55s}), the neutrino evolution is adiabatic in all channels for neutrino energies below $E\lesssim 1\;{\rm MeV}$ and adiabatic for $P^\text{(m)}_{\text{33}}$ up to $E\lesssim 5\;{\rm MeV}$. At higher energies the general trend is for the survival probabilities $P^\text{(m)}_{\text{11}}$, $P^\text{(m)}_{\text{22}}$ and $P^\text{(m)}_{\text{33}}$ to decrease with energy but in no channel at this epoch does the evolution become fully diabatic. 
\item at early times there is little line-of-sight dependence, 
\item at the same epoch and the same normal mass ordering the antineutrino transition probabilities (not shown) at all energies are close to adiabatic,
\item after the peak luminosity, Figure (\ref{fig:N100vSmNHAt1s}), the neutrino evolution remains adiabatic for the low energies $E\lesssim 1\;{\rm MeV}$. At higher energies the evolution along some lines of sight becomes more adiabatic - compare $P^\text{(m)}_{\text{33}}$ from Figure (\ref{fig:N100vSmNHAt0.55s}) with $P^\text{(m)}_{\text{33}}$ from Figure (\ref{fig:N100vSmNHAt1s}) - but along other lines of sight the evolution is similar to the earlier epoch,
\item the line-of-sight dependence is considerable with up to 40\% differences in the transition probabilities for some energies 
in certain channels,
\item at the same epoch and the same normal mass ordering the antineutrino transition probabilities during this period show only a small departure from adiabaticity in the $\bar{\nu}_{1} \leftrightarrow \bar{\nu}_2$ mixing channel at the level of 10\%. 
\item at late times, Figure (\ref{fig:N100vSmNHAt1.3s}) the neutrino evolution has become close to adiabatic in all mixing channels except for one line of sight. 
\item at late times the antineutrino transition probabilities for the normal mass ordering are adiabatic at all energies.
\end{itemize}

\begin{figure}[t!]
\includegraphics[trim={0 0 0 2cm},clip,width=1\linewidth]{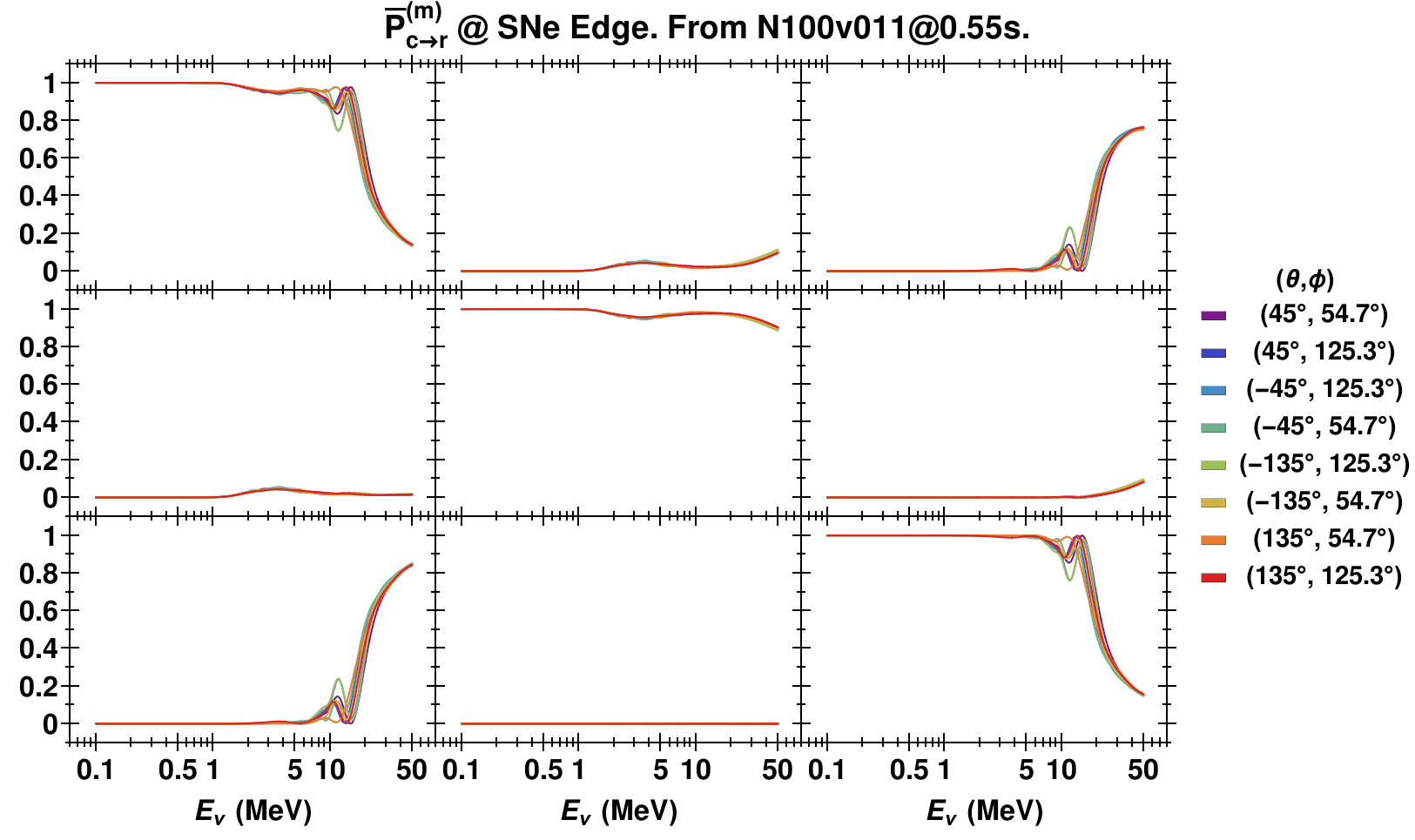}
\caption{The same as Figure (\ref{fig:N100vSmNHAt0.55s}) but for inverted mass ordering and antineutrinos.}
\label{fig:N100vSmBarIHAt0.55s}
\end{figure}
\begin{figure}[b!]
\includegraphics[trim={0 0 0 2cm},clip,width=1\linewidth]{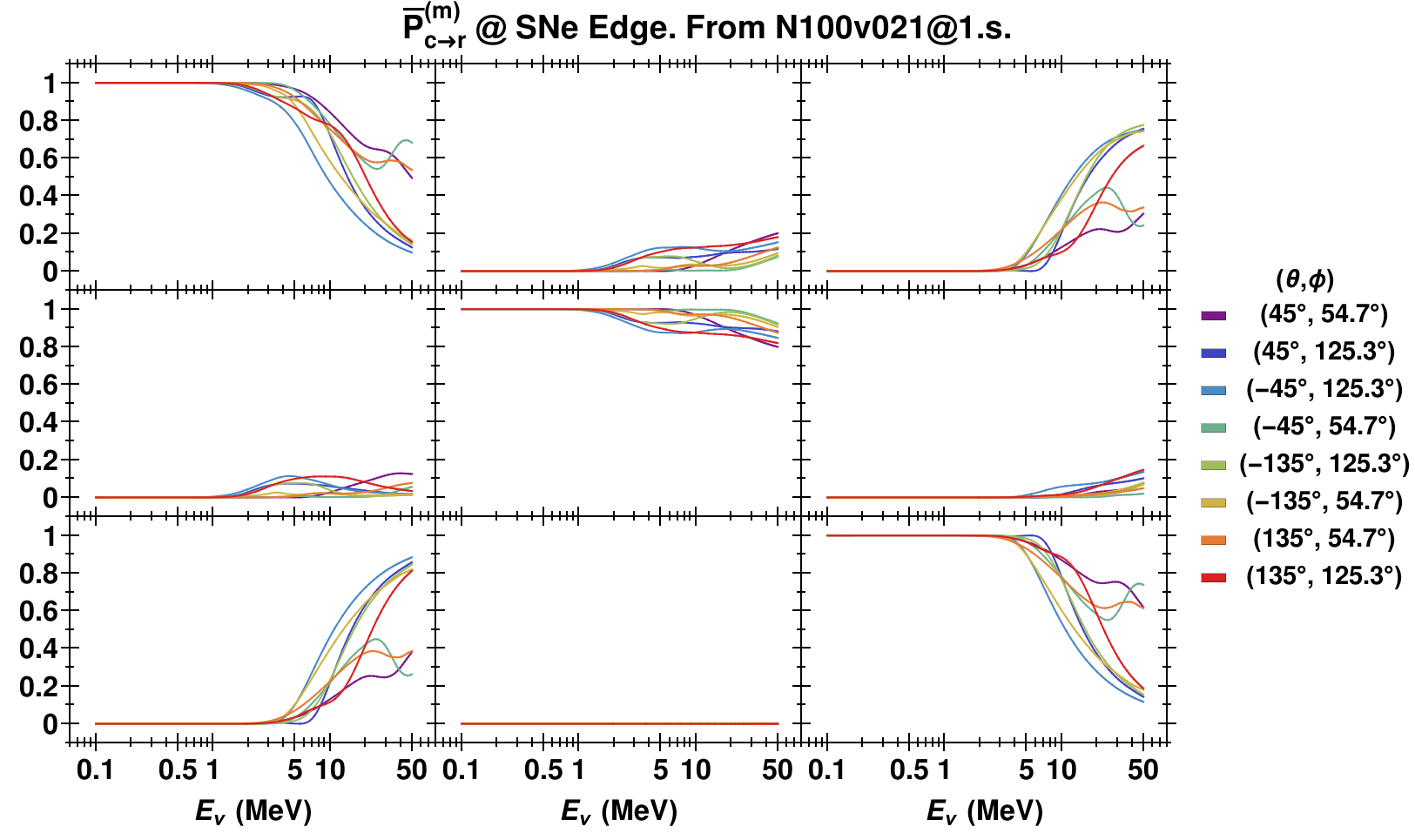}
\caption{The same as Figure (\ref{fig:N100vSmNHAt1s}) but for inverted mass ordering and antineutrinos.}
\label{fig:N100vSmBarIHAt1s}
\end{figure}
\begin{figure}[t!]
\includegraphics[trim={0 0 0 2cm},clip,width=1\linewidth]{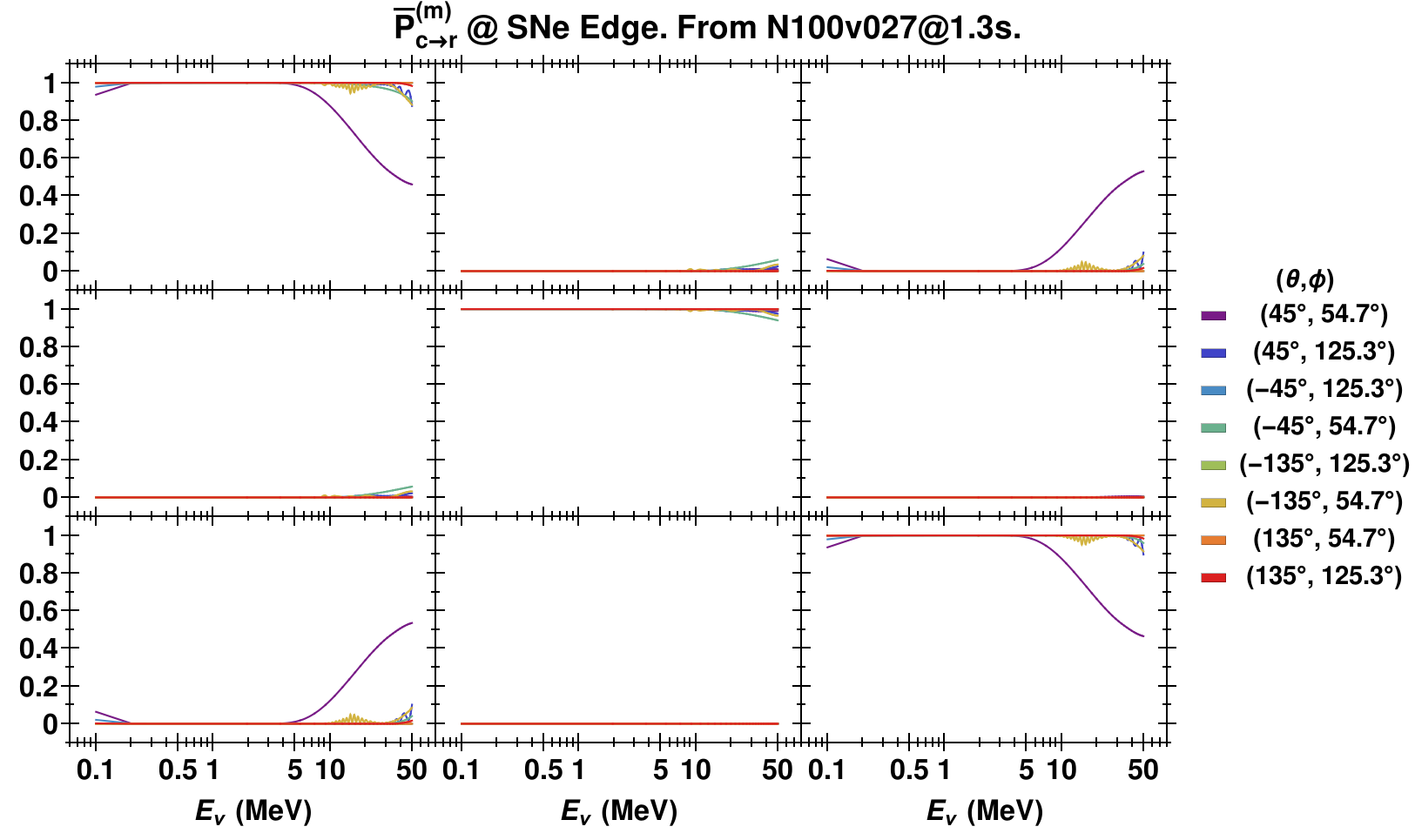}
\caption{The same as Figure (\ref{fig:N100vSmNHAt1.3s}) but for the inverted mass mass ordering and antineutrinos.}
\label{fig:N100vSmBarIHAt1.3s}
\end{figure}
For antineutrinos and an inverted mass ordering the results are very similar as a function of both time and energy.
\begin{itemize}
\item at early times, Figure (\ref{fig:N100vSmBarIHAt0.55s}), there is little line-of-sight dependence,
\item the evolution at this epoch is adiabatic for $P^\text{(m)}_{\text{11}}$ and $P^\text{(m)}_{\text{22}}$ up to $E\lesssim 1\;{\rm MeV}$ only, but up $E\lesssim 5\;{\rm MeV}$ for the antineutrinos $\bar{P}^\text{(m)}_{\text{11}}$ and $\bar{P}^\text{(m)}_{\text{33}}$. The transition probabilities $P^\text{(m)}_{\text{33}}$ and $\bar{P}^\text{(m)}_{\text{22}}$ are close to unity for all energies at this epoch.
\item midway through the signal, Figure (\ref{fig:N100vSmBarIHAt1s}), the line-of-sight dependence emerges for both neutrinos and antineutrinos with energies above $E \gtrsim 1\;{\rm MeV}$ with differences in the transition probabilities of order 50\% in $P^\text{(m)}_{\text{11}}$, $P^\text{(m)}_{\text{12}}$, $P^\text{(m)}_{\text{21}}$ and $P^\text{(m)}_{\text{22}}$, and the set $\bar{P}^\text{(m)}_{\text{11}}$, $\bar{P}^\text{(m)}_{\text{13}}$, $\bar{P}^\text{(m)}_{\text{31}}$ and $\bar{P}^\text{(m)}_{\text{33}}$.
\item at this epoch the transition probability $P^\text{(m)}_{\text{33}}$ remains close to unity at all energies but the survival probability $\bar{P}^\text{(m)}_{\text{22}}$ begins to exhibit some departure from adiabaticity for $E\gtrsim 1\;{\rm MeV}$ at the level of 20\%.
\item at late times, Figure (\ref{fig:N100vSmBarIHAt1.3s}), the line-of-sight dependence mostly disappears and the transition probabilities in all neutrino and all antineutrino channels becomes mostly adiabatic at all energies.
\end{itemize}

\begin{figure}[ht]
\includegraphics[width=1\linewidth]{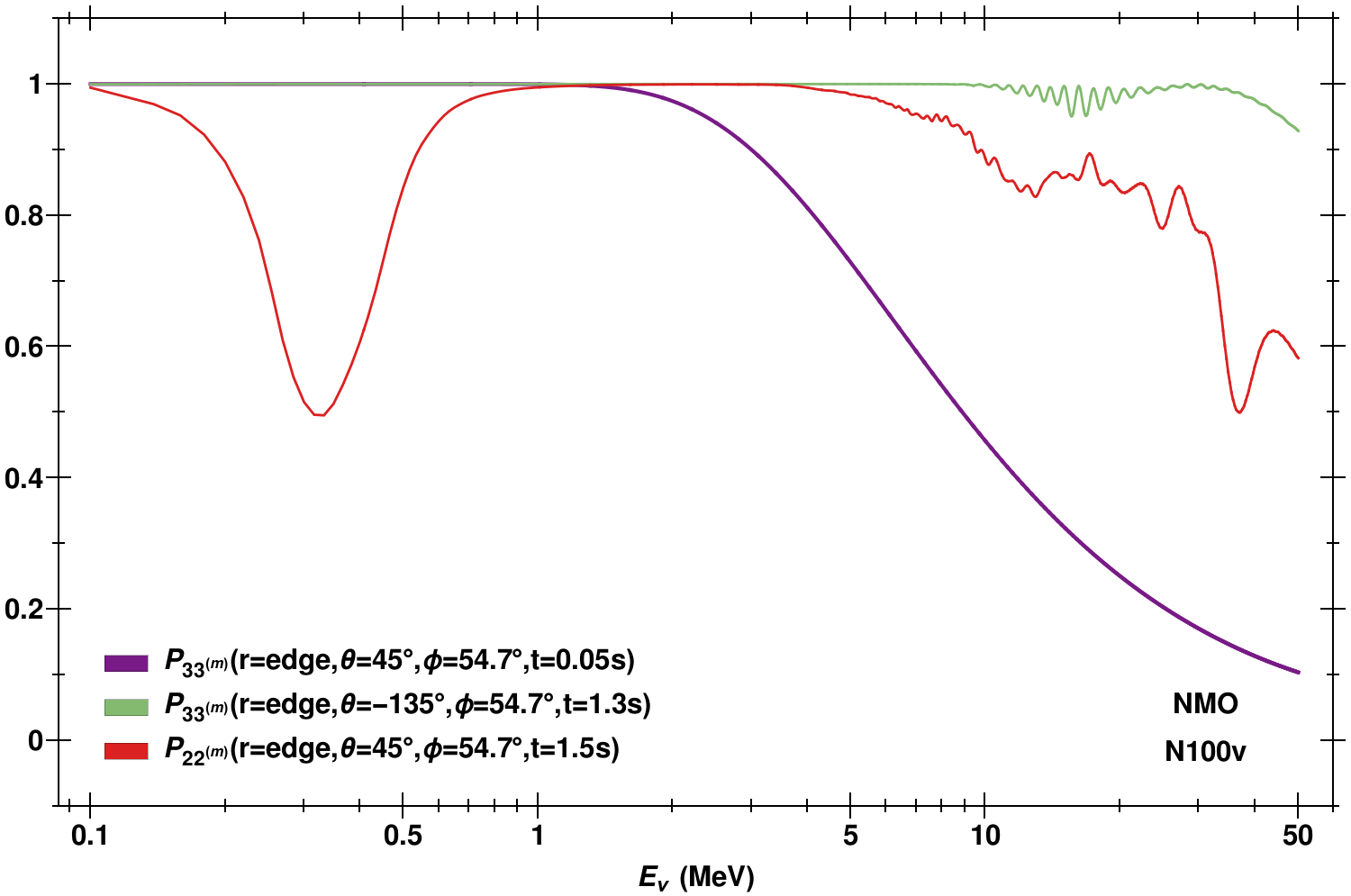}
\caption{$\nu^\text{(m)}$ survival probability vs energy for a particular choice zenith and azimuth angles.}
\label{fig:N100vSelectedSm}
\end{figure}
Before moving on to present the fluxes on Earth, we briefly consider how the line-of-sight dependence emerges. 
In Figure (\ref{fig:N100vSelectedSm}) we show three examples of the normal mass ordering matter basis transition probability as a function of energy for three different trajectory and snapshot time choices. These three specific probabilities were chosen because they serve as good examples of the rich oscillation phenomenology present in supernova environments. We notice in all three examples how the matter oscillation probability deviates from unity but, as we shall show, the diabatic evolution has a variety of causes.
\begin{figure}[ht]
	\centering
    \begin{minipage}{0.99\textwidth}
        \centering
        \includegraphics[width=1\linewidth]{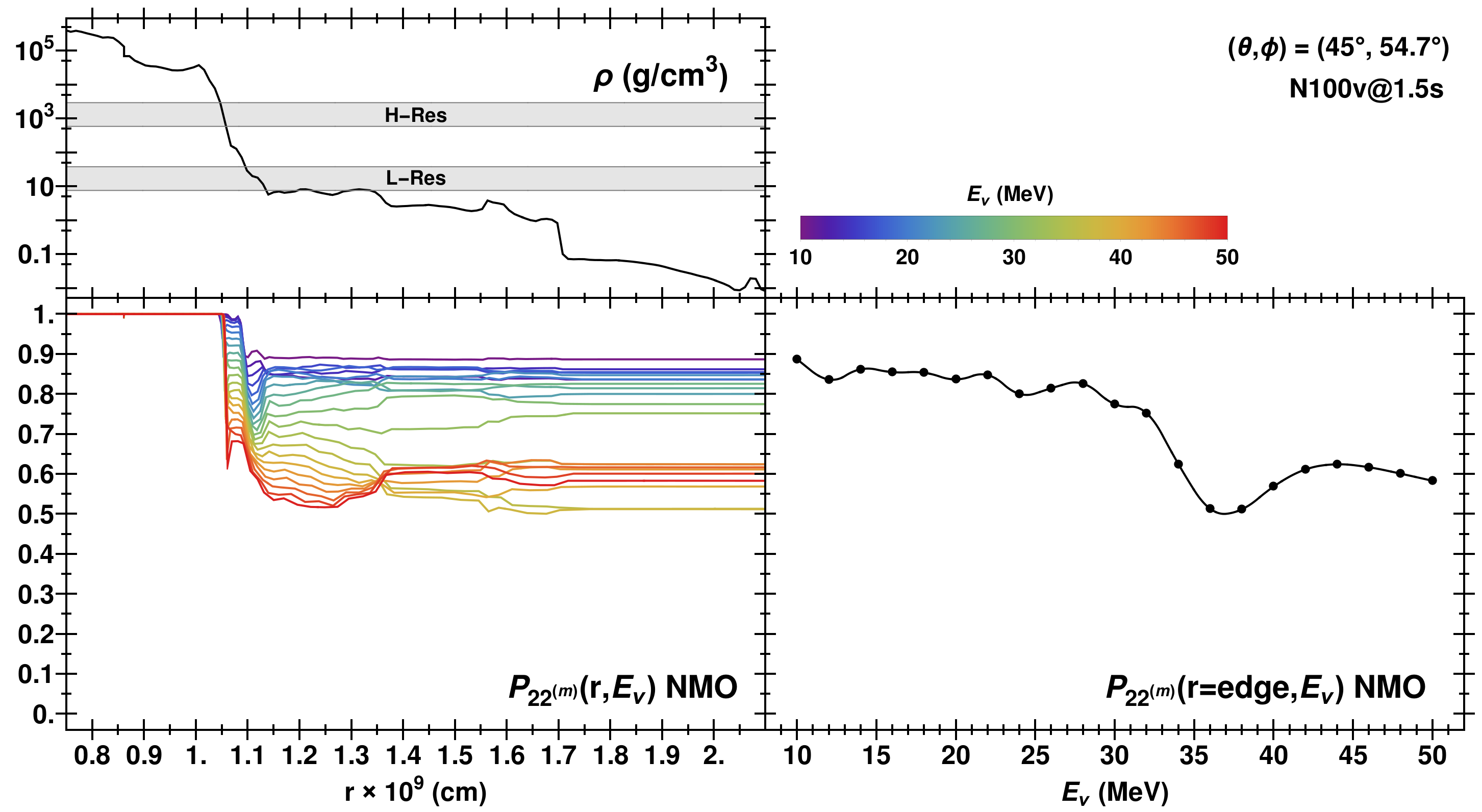}
    \end{minipage}
        \caption{Turbulent-like effects in the $\nu_2^\text{(m)}$ survival probability at $t=1.5\text{s}$. The upper left panel shows the density of the SN material along the neutrino trajectory defined by the zenith and azimuth angels given in the upper right corner. The gray "H-Res" and "L-Res" bands show the density range corresponding to the High and Low MSW resonances for the energy range given. The lower left panel shows how the $\nu_2^\text{(m)}$ survival probability changes with distance from the SN center for a range of energies. The lower right panel shows the $\nu_2^\text{(m)}$ survival probability as a function of neutrino energy as measured after the neutrino has traversed all of the SN material and is now essentially in vacuum. There are fewer points in the lower right panel than are in Figure (\ref{fig:N100vSelectedSm}); these points match the values in the lower left panel where too many points would make the plot unreadable. A normal neutrino mass ordering is assumed.}
        \label{fig:Turb}
\end{figure}
\begin{figure}[ht]
	\centering
    \begin{minipage}{0.99\textwidth}
        \centering
        \includegraphics[width=1\linewidth]{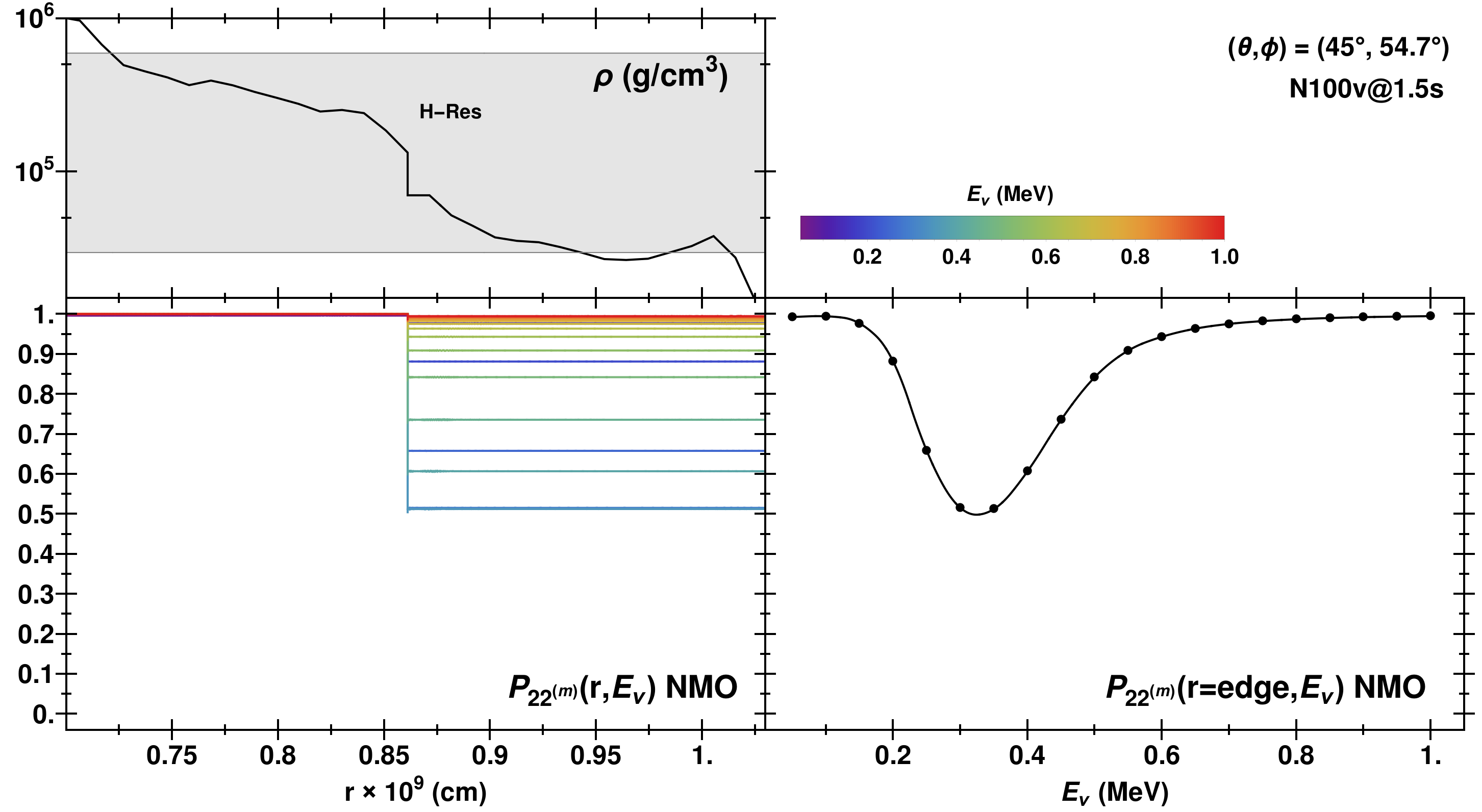}
    \end{minipage}
        \caption{The effect of discontinuous density on $\nu_2^\text{(m)}$ survival probability at $t=1.5\text{ s}$. The structure and layout of the figure is the same as that of Figure (\ref{fig:Turb}).}
        \label{fig:Disco}
\end{figure}
\begin{figure}[ht!!!!!]
	\centering
    \begin{minipage}{0.99\textwidth}
        \centering
        \includegraphics[width=1\linewidth]{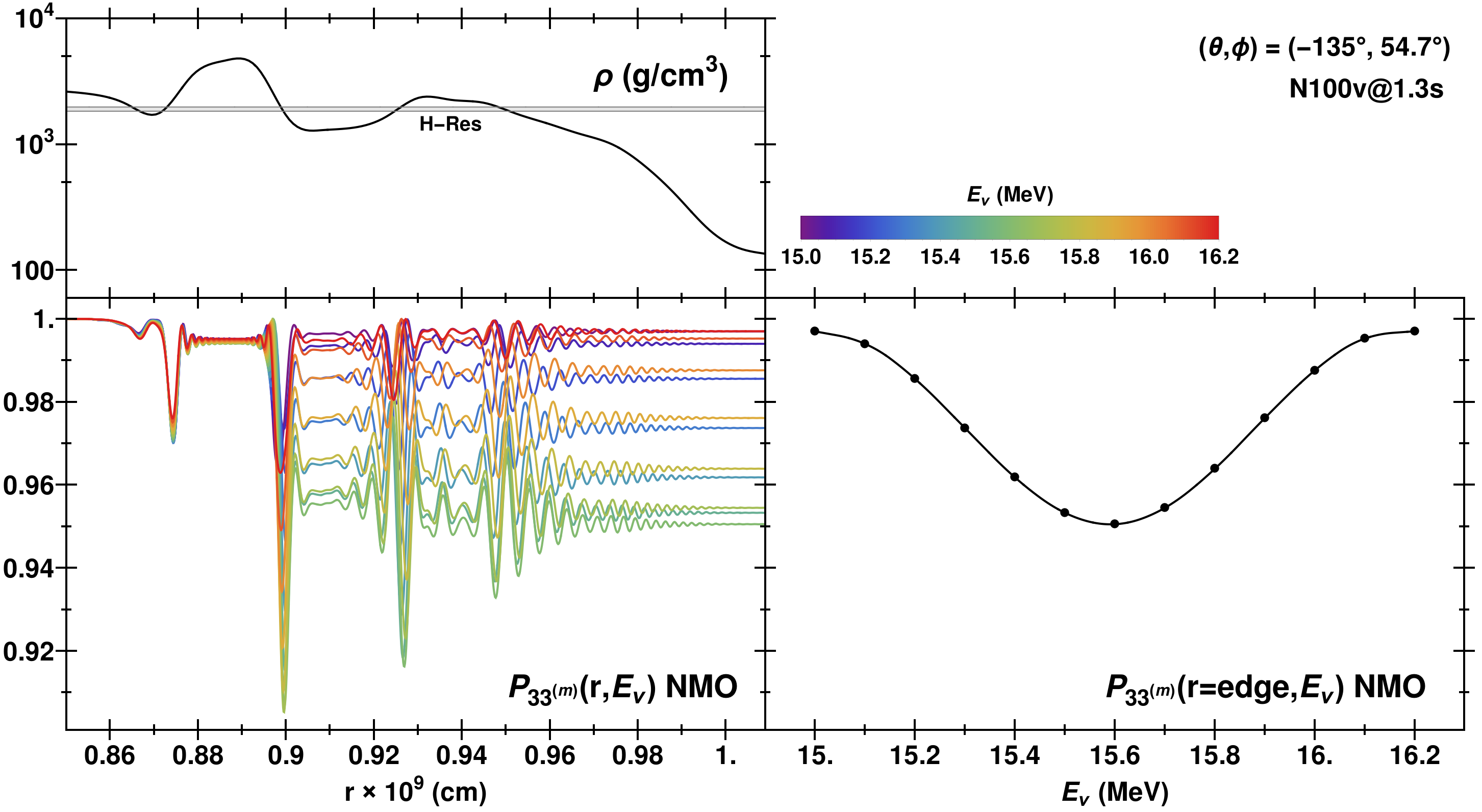}
    \end{minipage}
        \caption{Oscillatory effects in the $\nu_3^\text{(m)}$ survival probability at $t=1.3\text{ s}$. The structure and layout of the figure is the same as that of Figure (\ref{fig:Turb}).}
        \label{fig:Osc}
\end{figure}

\begin{itemize}
\item The most common oscillation effect is illustrated by the purple line in Figure (\ref{fig:N100vSelectedSm}) which corresponds to the $\nu_3^\text{(m)}$ survival probability early on in the SN process. The deviation from unity occurs at the star's edge, where the density plummets (as depicted in Figure (\ref{fig:RhoYeProfile})). The density profile drops rapidly through both the high and low Mikheyev-Smirnov-Wolfenstein (MSW) resonances and at approximately these values, diabatic effects turn on. These diabatic effects have an energy dependence of approximately $e^{-a/E_\nu}$ where $a$ is an energy independent factor that depends on the density profile and the vacuum mixing angles and masses \cite{Parke:1986jy}. This energy dependence explains why $\nu_3^\text{(m)}$ survives at low energy ($E_\nu<1\text{ MeV}$) but disappears for high energy ($E_\nu>10\text{ MeV}$). 
\item The red curve in Figure (\ref{fig:N100vSelectedSm}) shows the $\nu_2^\text{(m)}$ survival probability late in the SN process and has two interesting features.
\begin{enumerate}
\item The first feature is the high energy ($E_\nu>10\text{ MeV}$) region where the survival probability decreases somewhat randomly starting at $5\text{ MeV}$. Figure (\ref{fig:Turb}) shows how this $\nu_2^\text{(m)}$ survival probability changes as the neutrino propagates through the SN. This figure shows the diabatic effect of the density crossing the High MSW resonance (H-Res) (but before the L-Res). The results are an energy-dependent spread in survival probability which, if the L-Res was absent, would result in something similar to the purple line in Figure (\ref{fig:N100vSelectedSm}). But, unlike the H-Res case, the density profile does not plummet through the L-Res. Instead it stays near the L-Res for quite a while. This is because, by $t=1.5\text{s}$, the SN explosion has pushed some of the stellar material out to these distances. This material does not have a smooth density profile and every time it changes suddenly it causes diabatic effects because the density is near the L-Res. Thus some of the turbulence in the density imprints on the oscillation probability. The effects of turbulent density profiles on neutrino oscillation probability is an important field of research all by itself (see \cite{Patton2015a} and \cite{Mirizzi2015a} for reviews).
\item The second feature is the low energy ($E_\nu<1\text{ MeV}$) region where the survival probability dips around $0.3\text{ MeV}$. Figure (\ref{fig:Disco}) shows how this $\nu_2^\text{(m)}$ survival probability changes as the neutrino propagates through the SN. The density profile shows a discontinuous jump which corresponds to the edge of a deflagration flame. This density discontinuity in the H-Res density range produces an energy dependent dip in the survival probability. If the discontinuity were removed the oscillation probability would not be affected at all. The importance of density discontinuities on neutrino oscillation physics has been previously considered the context of core-collapse SNe \cite{Lund2013} where the shock fronts cause the density discontinuities.
\end{enumerate}
\item The green curve in Figure (\ref{fig:N100vSelectedSm}) shows the $\nu_3^\text{(m)}$ survival probability late in the SN process and has an interesting oscillatory feature. Figure (\ref{fig:Osc}) shows how this $\nu_3^\text{(m)}$ survival probability changes as the neutrino propagates through the SN. This figure reveals that the neutrino passes through a region where the density steeply fluctuates around the H-Res. This imprints an oscillatory feature on the survival probability that looks very similar to the phenomenon of `phase-effects' \cite{2006PhRvD..73e6003K,2007PhRvD..75i3002D}.
\end{itemize}


\section{Neutrino Fluxes on Earth}

\begin{figure}[ht]
	\centering
    \begin{minipage}{0.99\textwidth}
        \centering
        \includegraphics[trim={0 0 0 2.2cm},clip,width=1\linewidth]{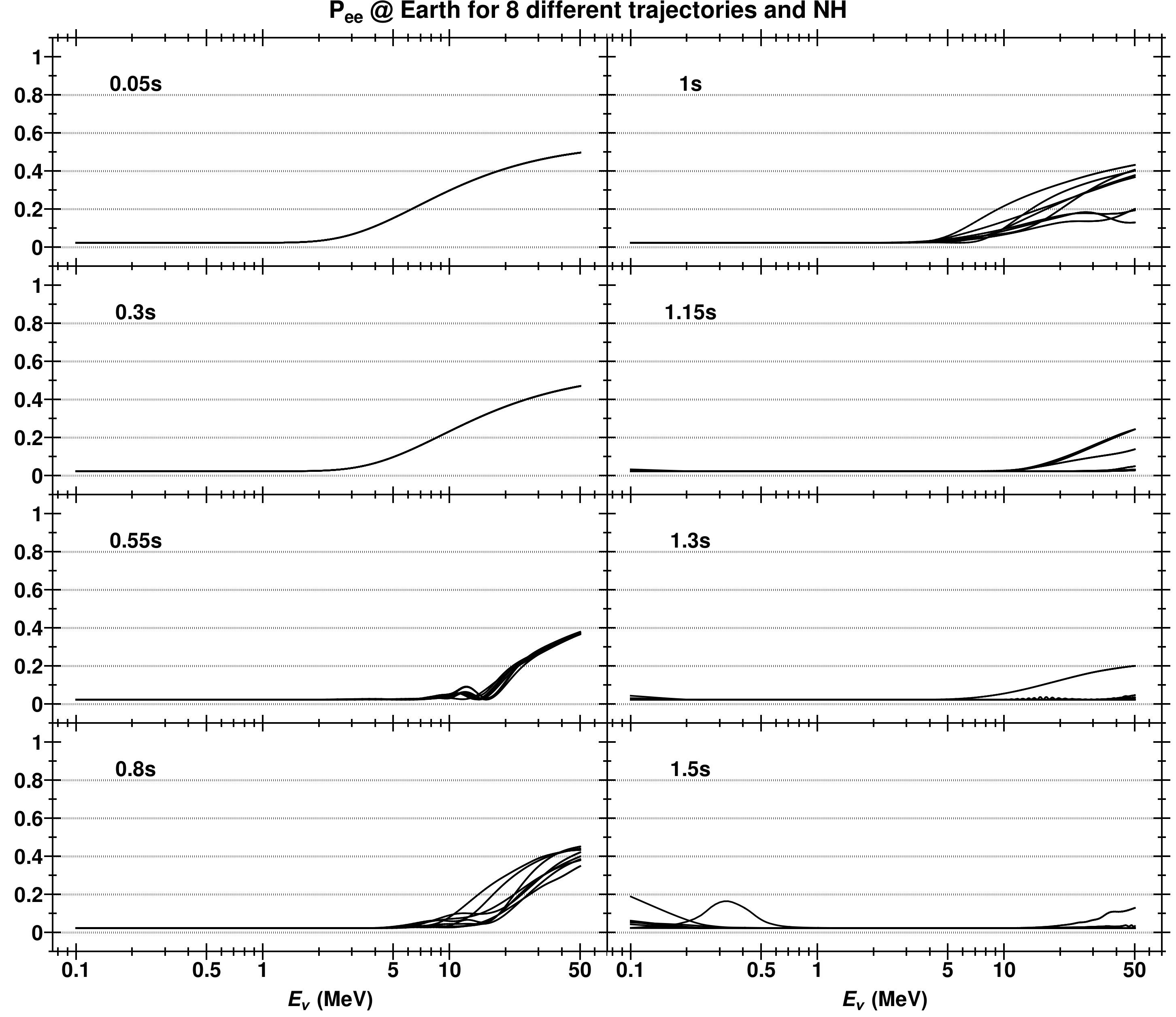}
    \end{minipage}
        \caption{$P_{ee}$ on Earth for different times and trajectories and NMO.}
        \label{fig:N100vPeeEarthTblNH}
\end{figure}

\begin{figure}[ht]
	\centering
    \begin{minipage}{0.99\textwidth}
        \centering
        \includegraphics[trim={0 0 0 2.6cm},clip,width=1\linewidth]{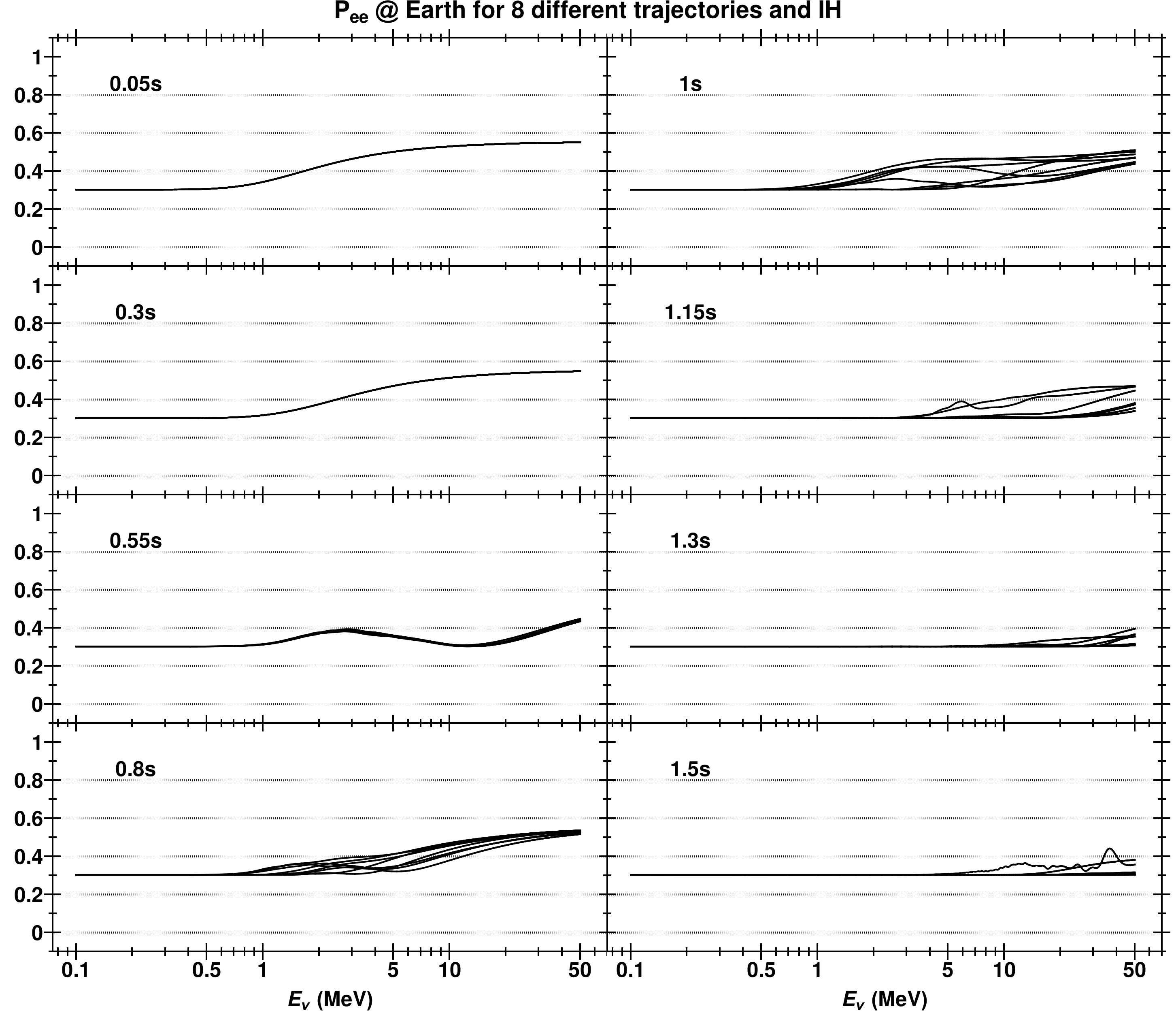}
    \end{minipage}
        \caption{$P_{ee}$ on Earth for different times and trajectories and IMO.}
        \label{fig:N100vPeeEarthTblIH}
\end{figure}

\begin{figure}[ht]
	\centering
    \begin{minipage}{0.99\textwidth}
        \centering
        \includegraphics[trim={0 0 0 2.6cm},clip,width=1\linewidth]{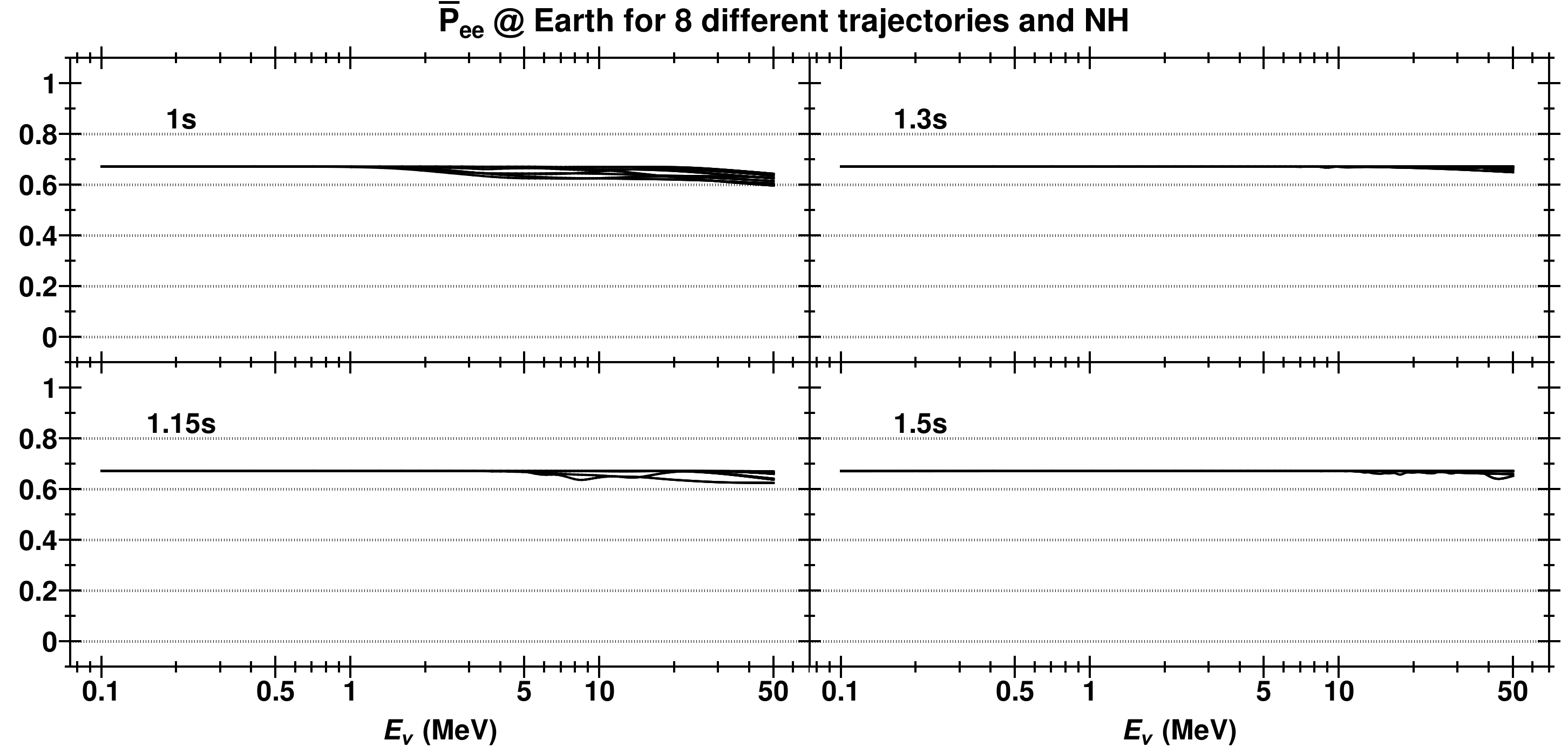}
    \end{minipage}
        \caption{$\bar{P}_{ee}$ on Earth for different times and trajectories and NMO.}
        \label{fig:N100vPBareeEarthTblNHLateTimes}
\end{figure}

\begin{figure}[ht]
	\centering
    \begin{minipage}{0.99\textwidth}
        \centering
        \includegraphics[trim={0 0 0 2.6cm},clip,width=1\linewidth]{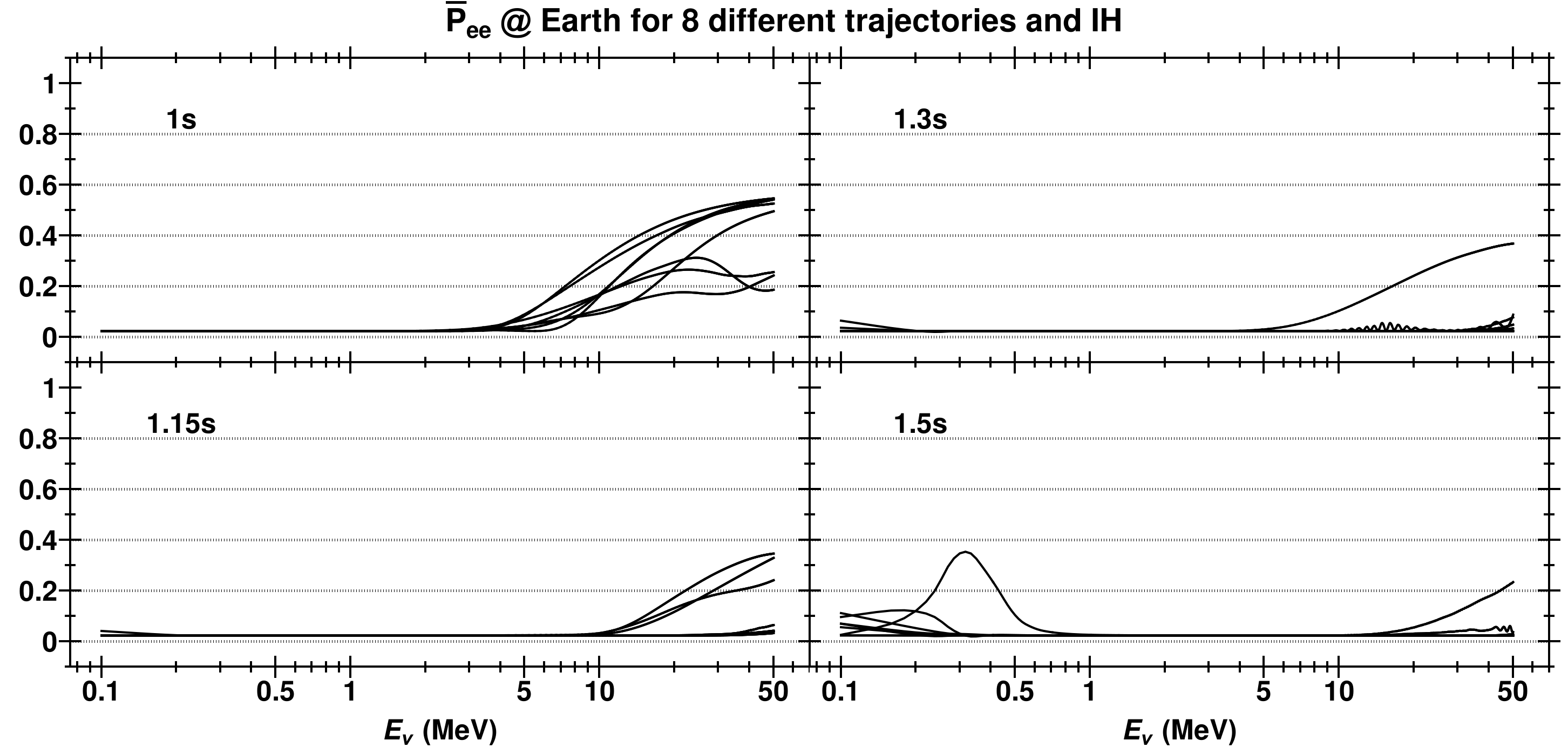}
    \end{minipage}
        \caption{$\bar{P}_{ee}$ on Earth for different times and trajectories and IMO.}
        \label{fig:N100vPBareeEarthTblIHLateTimes}
\end{figure}

\begin{figure}[ht!]
	\centering
    \begin{minipage}{0.8\textwidth}
        \centering
        \includegraphics[trim={0 0 0 1.4cm},clip,width=1\linewidth]{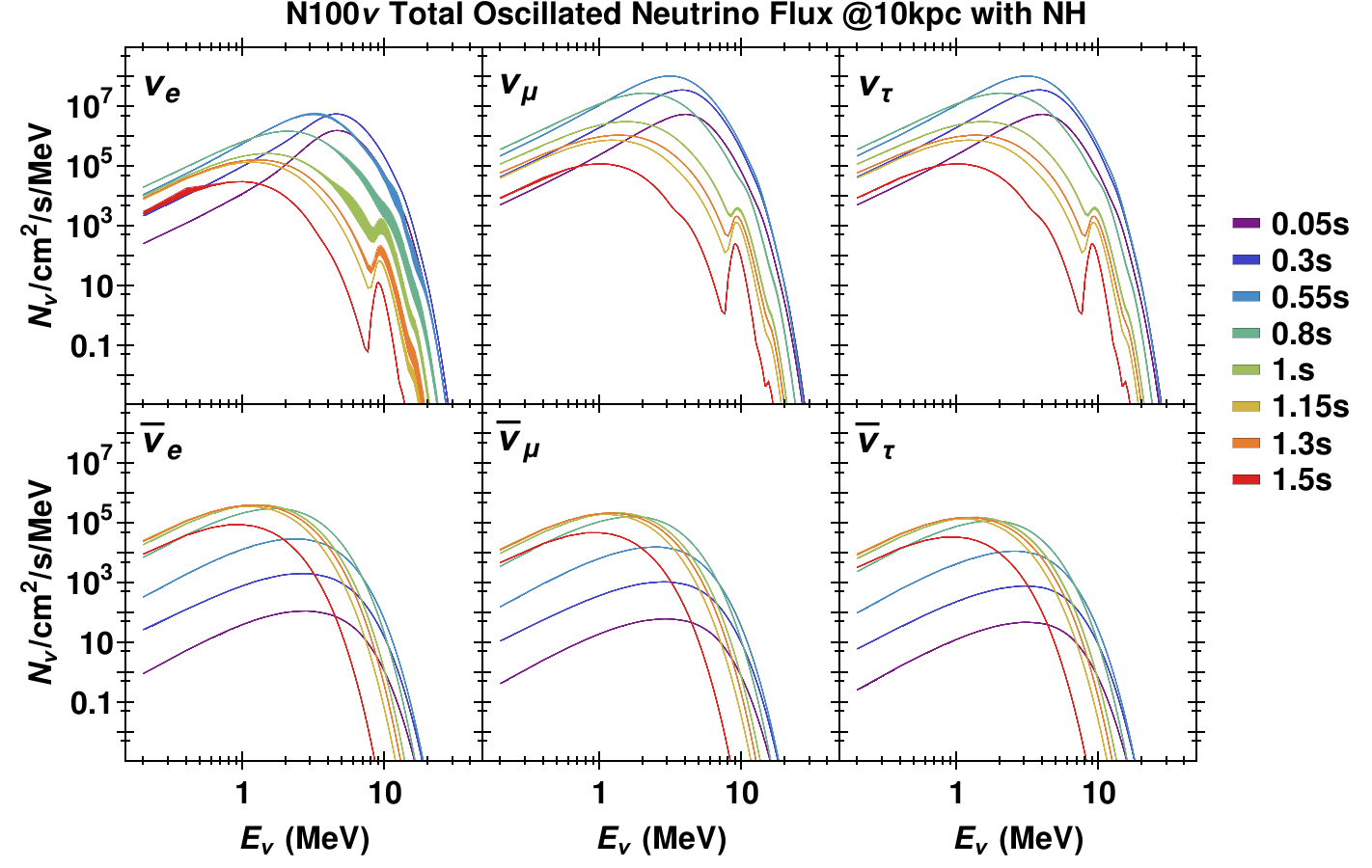}
    \end{minipage}
    \begin{minipage}{0.8\textwidth}
    	\vspace{5 mm}
        \centering
        \includegraphics[trim={0 0 0 1.4cm},clip,width=1\linewidth]{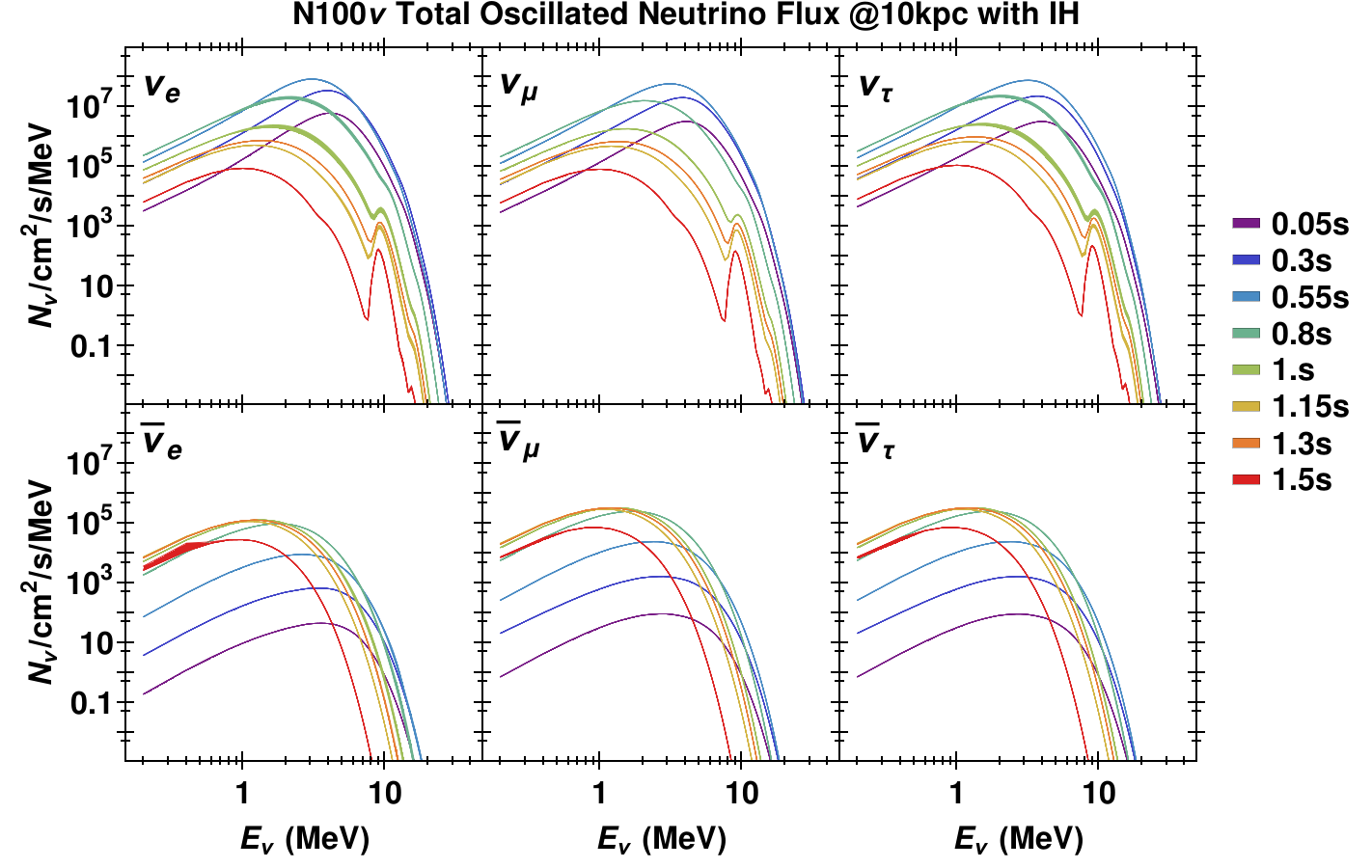}
    \end{minipage}
        \caption{The total oscillated neutrino flux at 10 kpc for the N100$\nu$ model. The top figure has normal mass ordering and the bottom figure has inverted mass ordering.}
        \label{fig:OscFlux}
\end{figure}
\begin{figure}[ht]
	\centering
    \begin{minipage}{0.8\textwidth}
        \centering
        \includegraphics[trim={0 0 0 1.4cm},clip,width=1\linewidth]{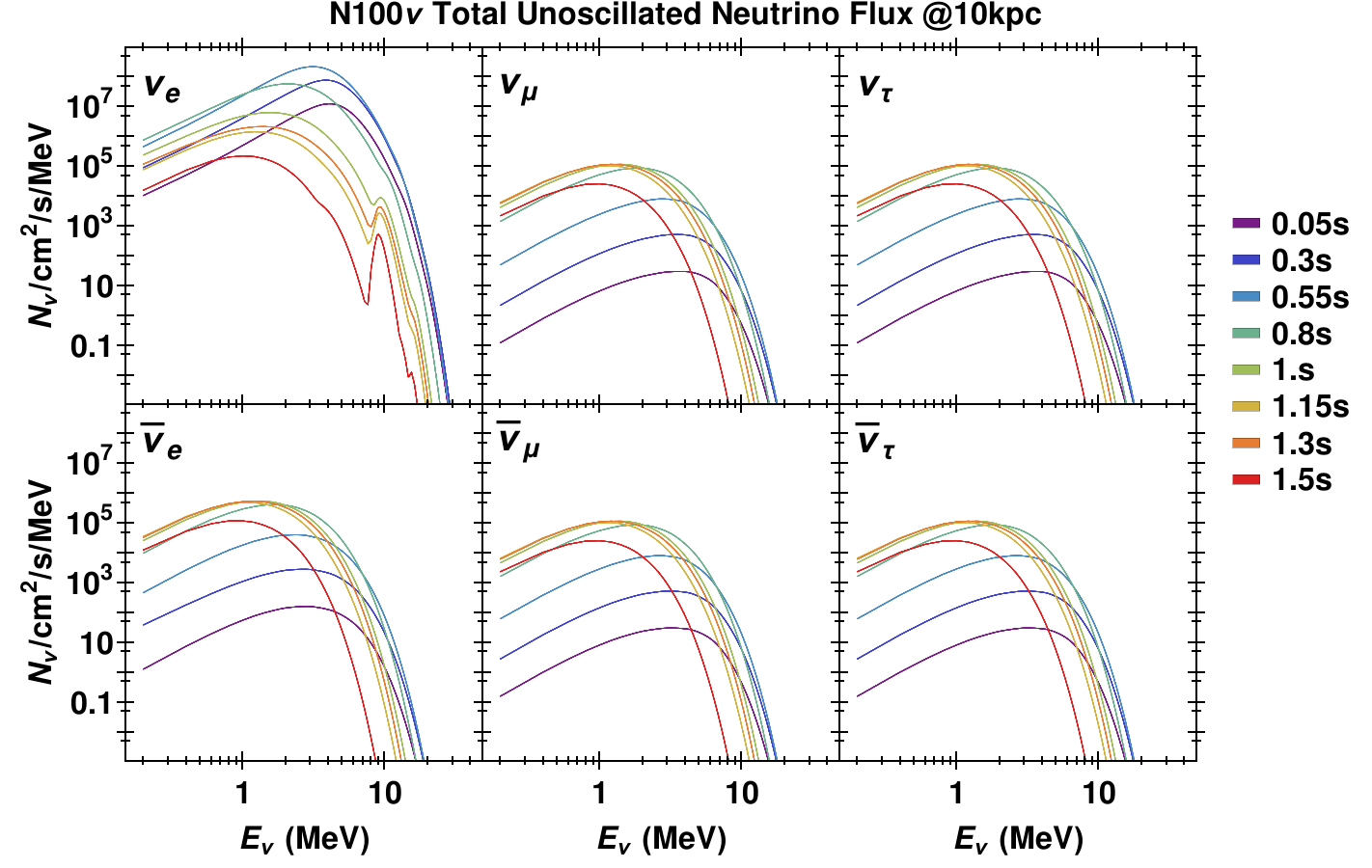}
    \end{minipage}
        \caption{The total unoscillated neutrino flux at 10 kpc for the N100$\nu$ model.}
        \label{fig:UnoscFlux}
\end{figure}

Now we have computed the transition probabilities through the supernovae, we now fold in the the vacuum decoherence, discussed at the end of Section \S\ref{subsec:Theory} in Equation (\ref{eqn:Decoherence}). Figures (\ref{fig:N100vPeeEarthTblNH}) and (\ref{fig:N100vPeeEarthTblIH}) show the $\nu_e$ survival probability on Earth for the normal and inverted mass ordering respectively as a function of neutrino energy for each of the 8 time slices considered in this paper. In each of the subplots at the different time slices, there are eight lines representing the 8 different angular trajectories under consideration. Figures (\ref{fig:N100vPBareeEarthTblNHLateTimes}) and (\ref{fig:N100vPBareeEarthTblIHLateTimes}) show the survival probability for the antineutrinos in the two orderings but only for the last four time slices. 
Careful examination of the figures reveals that all of the oscillation phenomena discussed above are visible in these probabilities. Another obvious feature is the emergence of the line-of-sight dependence at $t\sim 0.6\;{\rm s}$ and its later disappearance at $t\sim 1.2\;{\rm s}$. Decoherence reduces the amount of the line-of-sight dependence to roughly 30\% at its peak when t=1 s. How these variations in probability affect the measurable flux will be discussed in the next section. 

To calculate the flux on Earth the oscillation probabilities need to be multiplied by the source flux. Given a source flux $\Phi^\text{(s)}=\left(\Phi_e^\text{(s)},\Phi_\mu^\text{(s)},\Phi_\tau^\text{(s)}\right)^T$ the flux on Earth $\Phi^\text{(e)}=\left(\Phi_e^\text{(e)},\Phi_\mu^\text{(e)},\Phi_\tau^\text{(e)}\right)^T$ can be given by
\begin{align}
	\Phi^\text{(e)}=\frac{1}{4\pi R^2}\,P\,\Phi^\text{(s)}\label{eqn:OscFlux},
\end{align}
where $R$ is the distance from Earth to the SN and $P$ is the matrix of probabilities with elements given by Equation (\ref{eqn:Decoherence}). Figure (\ref{fig:OscFlux}) shows the results of Equation (\ref{eqn:OscFlux}) for each of the 3 neutrino and anti-neutrino flavors using a supernova distance of 10 kpc. The different colored lines show how the flux changes with time and the width of each line represents the variation in the flux across the eight trajectories considered. For reference, Figure (\ref{fig:UnoscFlux}) displays the same unoscillated flux as Figure (\ref{fig:N100vTotalNeutrinoLuminosity}) but at 10 kpc and in the same units as the flux displayed in Figure (\ref{fig:OscFlux}). Figures (\ref{fig:OscFlux} \& \ref{fig:UnoscFlux}) show that, for both NMO and IMO, the $\nu_e$ flux emitted in the core of the supernova has mostly been transformed into a flux of $\nu_\mu$ and $\nu_\tau$. This effect is more pronounced in NMO than in IMO. Additionally, this effect also means that the 10 MeV peak in the unoscillated $\nu_e$ flux becomes imprinted on the $\nu_\mu$ and $\nu_\tau$ oscillated fluxes. Finally, compared to the overall flux shape and the difference between successive time slices, the variability due to line-of-sight dependence is a subordinate effect. 

The oscillated fluxes depicted in Figure (\ref{fig:OscFlux}) need to be convolved with the neutrino cross-section and detector specifics in order to determine which, if any, of the neutrino production or oscillation features are observable at current or future neutrino detectors.


\section{Neutrino Detection \label{sec:NeutrinoDetection}}

\subsection{Detector Signals}
We now reach the final topic for discussion, namely the detectors, their event rates, and the sensitivity to the various features in the spectrum at the source and the features imprinted through the mantle of the supernova. We shall consider five different detectors listed in Table \ref{table:Detectors}, along with their detector mass and material. IceCube refers to just the main detector, not the DeepCore nor proposed PINGU subdetectors. These five detectors are representative of current and next-generation detectors. The event rates in detectors similar to Super-K, Hyper-K, JUNO, and DUNE can be calculated using \textsc{SNOwGLoBES} \cite{SNOwGLoBES}. \textsc{SNOwGLoBES} estimates detected event rates for relevant channels in the few to 100-MeV neutrino-energy range by folding neutrino flux with cross sections, and then applying a transfer matrix that takes into account both the distribution of interaction products for a given neutrino energy and the detector resolution effects. \textsc{SNOwGLoBES} outputs event distributions as a function of true neutrino energy, as well as realistic "smeared" distributions as a function of observed energy, with binning approximately matching typical detector resolutions.  The detector configurations used are denoted Super-K-"like", etc. because \textsc{SNOwGLoBES} provides representative transfer functions which approximate the detector response; detailed detector simulations are not publicly available for the detectors under consideration. We sum all interaction channels available in \textsc{SNOwGLoBES} for each detector. The individual interaction channel event rates are presented in Appendix \ref{Appendix:Channel} for Hyper-K and DUNE. The IceCube detector will be treated separately. 

\renewcommand{\arraystretch}{0.75}
\begin{table}[ht!!!]\centering
		\begin{tabular}{| l | c | c |}
			\cline{1-3}
			{\bf Detector}&{\bf Type}&{\bf Mass (kt)}\\ 
			\cline{1-3}
			Super-Kamiokande like: 30\% phototube coverage& Water Cherenkov&50\\  
			Hyper-Kamiokande like & Water Cherenkov& 560\\ 
			DUNE like detector&Liquid Ar&40\\ 
			JUNO like detector&Scintillator&20\\
   			IceCube & Water Cherenkov&3500$^*$\\            
			\cline{1-3}
		\end{tabular}
   \caption{Summary of the detectors under consideration.  Note that event rates simply scale by mass. $^*$For IceCube, the mass given is the effective mass used for the event rate calculation (see Appendix~\ref{Appendix:IceCube}).}
	\label{table:Detectors}
\end{table}
\renewcommand{\arraystretch}{1}
   

\subsection{Results}
\subsubsection{Super-K, Hyper-K, JUNO and DUNE}
\renewcommand{\arraystretch}{0.75}
\begin{table}\centering
		\begin{tabular}{| l | c | c | c |}
			\cline{1-4}
			{\bf Detector}&{\bf NMO}&{\bf IMO}&{\bf Unoscillated}\\ 
			\cline{1-4}
			Super-Kamiokande&$0.034 $&$0.076 $&$0.154$\\ 
			Hyper-Kamiokande&$0.378 $&$0.868 $&$1.725$\\ 
			DUNE&$            0.025 $&$0.066 $&$0.138$\\ 
			JUNO&$            0.014 $&$0.032 $&$0.063$\\
			IceCube*&$        0.286 $&$0.660 $&$1.320$\\
			\cline{1-4}  
		\end{tabular}
   \caption{Numbers of interactions per detector for each mass ordering and a SN at 10 kpc. These event counts are for the whole 1.5 s neutrino burst averaged over the 8 lines of sight considered. 
   The fourth column represents the number of interactions observed when neutrino oscillations are not taken into account. \newline * Note that the numbers of interactions quoted for IceCube are after background subtraction.}
	\label{table:Events}
\end{table}

\renewcommand{\arraystretch}{1}
In Table \ref{table:Events} we show the expected numbers of interactions for both mass orderings from all channels, energies, and for the full duration of the neutrino signal (1.5 s). These numbers are for a supernova distance of 10 kpc. Note that these are based only on interaction event rates, and the heretofore ignored detector efficiencies and energy smearing will decrease the chances of SNe Ia neutrino observation. 
The variation due to the line of sight we find to be small, $\sim 0.2\%$ for the normal mass ordering and  $\sim 0.6\%$ for the inverted. The low variations due to line of sight is a promising result because it means that all lines of sight contain the production and oscillation features of interest and therefore the feature detection probability is not decreased due to needing a particular line of sight.
\begin{figure}[ht!]
	\centering
    \begin{minipage}{0.99\textwidth}
        \centering
        \includegraphics[trim={0 0 0 1.3cm},clip,width=1\linewidth]{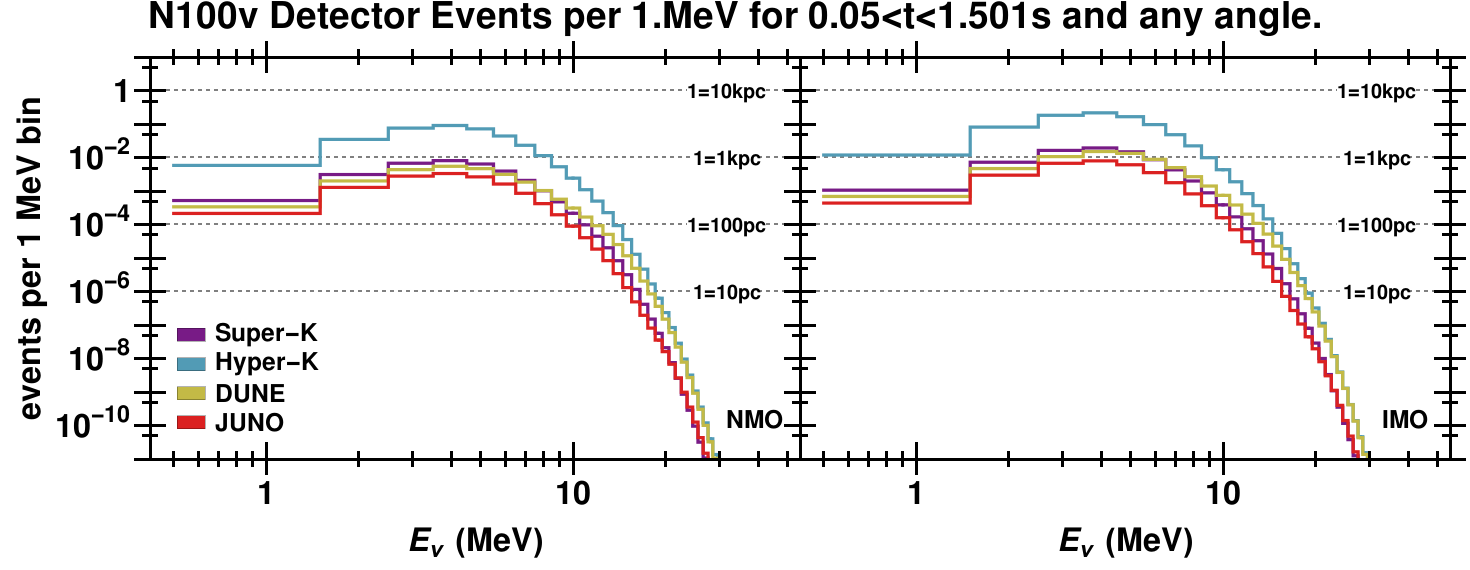}
    \end{minipage}
        \caption{All-channel neutrino interaction counts in 1-MeV bins for a SN Ia at 10 kpc. At the displayed scale, the line-of-sight angle has little effect and the plots are for NMO (left panel) and IMO (right panel). The different colored lines represent the interaction event numbers at the corresponding detector indicated in the legend. The horizontal lines are labeled to indicate how the vertical axis would shift for closer SNe. The plot represents the neutrino event count for the entire $\sim$1.5 s neutrino burst.}
        \label{fig:Events}
\end{figure}

As already known, SNe Ia are much dimmer neutrino sources than core-collapse supernovae to the extent that a type Ia supernova at 10 kpc will produce, at best, a few events assuming an upward few-$\sigma$ Poisson fluctuation. One would need a significantly closer SN Ia if one is to detect enough neutrinos to begin to observe discriminating features. In what follows we shall also consider placing the supernova at progressively closer distances of 1 kpc, 100 pc, and 10 pc. For each order-of-magnitude decrease in distance, the event rates increase by two orders of magnitude. Thus for a supernova at 5 kpc we expect a few events in Hyper-Kamiokande, but not until we decrease the distance to 1 kpc should we expect a few events in JUNO, DUNE, and Super-Kamiokande. At the same time, after fitting to the cumulative probability distance distribution in Adams \emph{et al.} \cite{2013ApJ...778..164A}, we find the probability that the next Galactic type Ia supernova is within this distance also decreases as approximately $\propto 1/d^{2.5}$. The chance that the next type Ia supernova is within $d=5\;{\rm kpc}$ is $\sim 10\%$ \cite{2013ApJ...778..164A} and within $d=1\;{\rm kpc}$ it is only $\sim 0.2\%$. 

\begin{figure}[ht!]
	\centering
    \begin{minipage}{0.99\textwidth}
        \centering
        \includegraphics[trim={0 0 0 1.3cm},clip,width=1\linewidth]{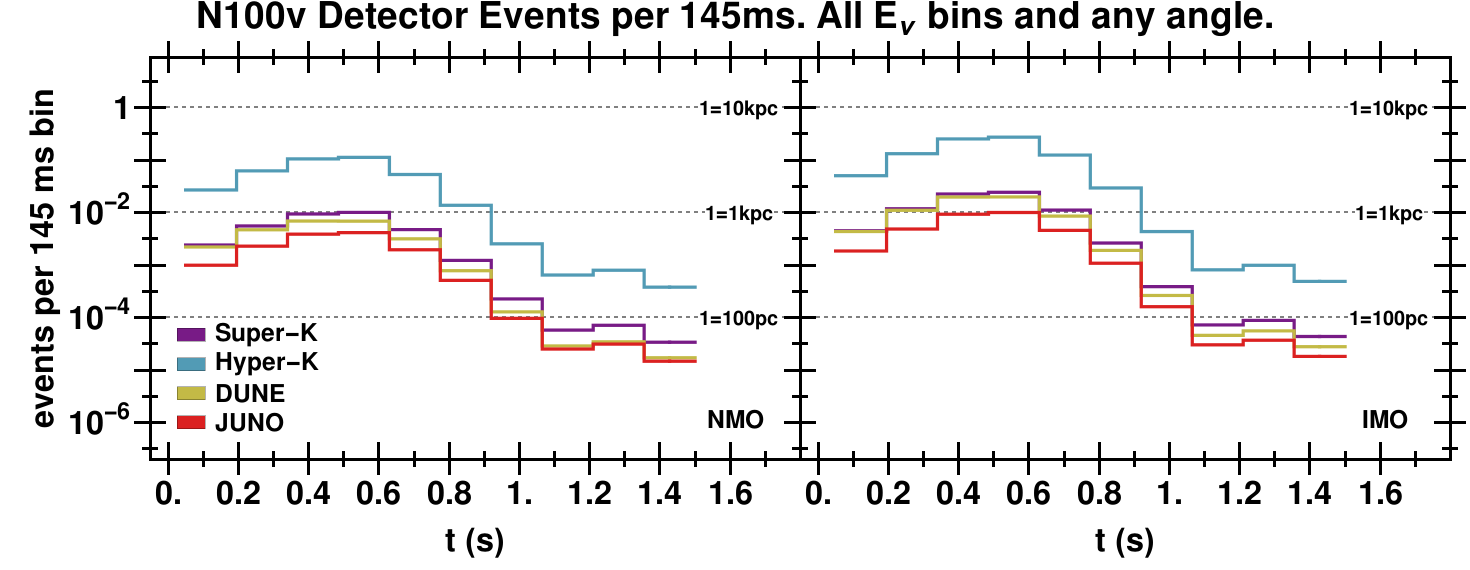}
    \end{minipage}
        \caption{All-channel, all-energy neutrino interaction rates per 145 ms for a SN Ia at 10 kpc. The plots are for a particular line-of-sight angle; however, at the displayed scale, the line-of-sight angle has little effect. The different colored lines represent the interaction event numbers at the corresponding detector indicated in the legend. The horizontal lines are labeled to indicate how the vertical axis would shift for closer SNe. The left plot is for NMO and the right plot for IMO.}
        \label{fig:EventsVsTime}
\end{figure}

The other factor affecting the event rates is the detector mass. As Table \ref{table:Events} indicates, the several hundred kiloton detector mass of Hyper-K brings Type Ia supernovae at close to the most probable distance of $d\sim10\;{\rm kpc}$ almost within reach. Another order-of-magnitude increase in detector mass would bring virtually every Type Ia supernova in the Galaxy within the scope of a detector, another benefit of the 5 mT water Cherenkov detector discussed by Kistler \emph{et al.} \cite{2011PhRvD..83l3008K}.

While event rates by themselves do have some discriminatory power, much more information is present in the spectrum should the supernova be close enough or the detector mass be sufficiently large that statistics permit partitioning. Figure (\ref{fig:Events}) displays the number of interaction events in 1-MeV bins expected in these same four detectors for a N100$\nu$ SN Ia again at 10 kpc. The events are summed over all interaction channels but, again, do not include detector efficiencies and energy smearing. The events are for the entire 1.5 s of the neutrino signal. As before, it is clear that at 10 kpc, it is doubtful that any neutrino signal will be detected by these next-generation neutrino detectors, but for a SN Ia at 1 kpc, the left plot in Figure (\ref{fig:Events}) reveals that Hyper-K will have a good chance of seeing the peak of the neutrino spectrum at $t\sim 0.6\;{\rm s}$. For the other detectors, the distance to the supernova would need to drop to 100 pc before they are able to see a significant portion of the neutrino spectrum. 

Figure (\ref{fig:EventsVsTime}) shows the all-channel, all-energy neutrino interaction rates in 145 ms bins for a SN Ia at 10 kpc. The event rates in all four detectors are seen to peak at $t \sim 0.5\;{\rm s}$ and drop off rapidly after $t \sim 1\;{\rm s}$.

\begin{figure}[ht!]
	\centering
    \begin{minipage}{0.99\textwidth}
        \centering
        \includegraphics[trim={0 0 0 1.3cm},clip,width=1\linewidth]{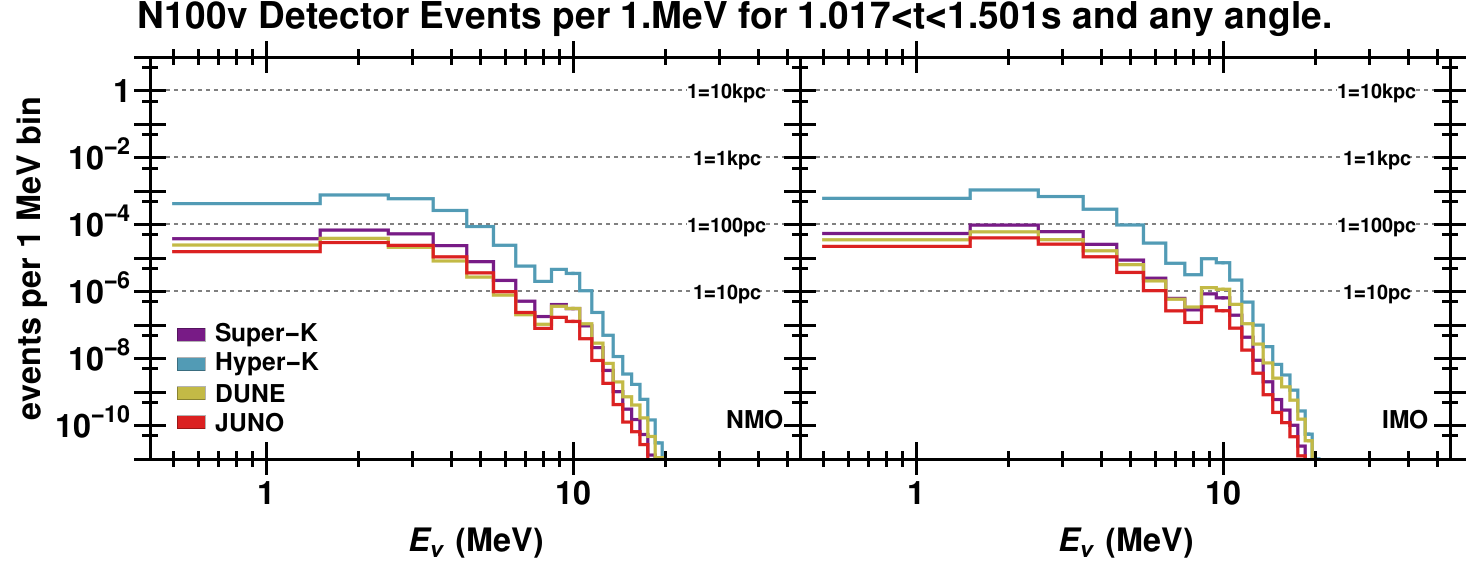}
    \end{minipage}
        \caption{All-channel neutrino event numbers in 1-MeV bins for a SN Ia at 10 kpc. At the displayed scale the line-of-sight angle has little effect and the plots are for NMO (left) and IMO (right). The different colored lines represent the interaction event rates at the corresponding detector indicated in the legend. The horizontal lines are labeled to indicate how the vertical axis would shift for closer SNe. The plot represents only the last 0.5 s of the neutrino burst where the 10-MeV emission from the iron group nuclei becomes apparent.}
        \label{fig:Events3of3}
\end{figure}
\begin{figure}[ht!]
	\centering
    \begin{minipage}{0.99\textwidth}
        \centering
        \includegraphics[trim={0 0 0 1.3cm},clip,width=1\linewidth]{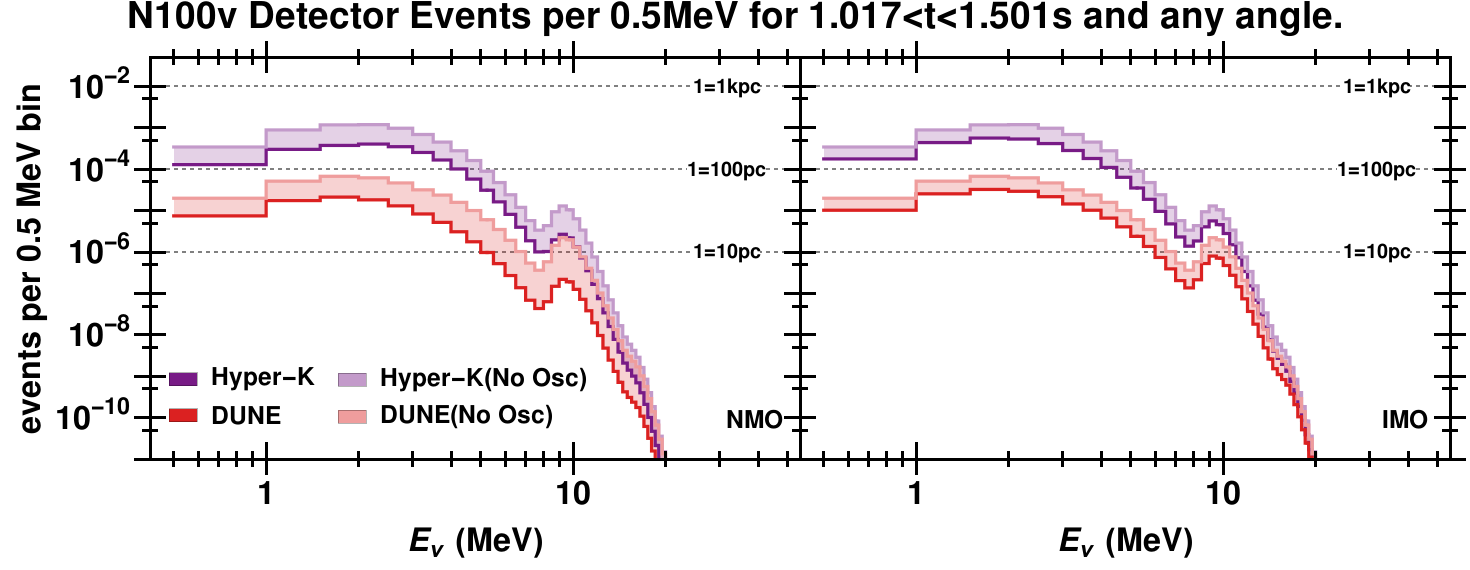}
    \end{minipage}
        \caption{Same as Figure (\ref{fig:Events3of3}) but including the detection rate assuming no neutrino oscillations and using 0.5-MeV bins.}
        \label{fig:Events3of3NoOscComp}
\end{figure}
The next significant feature in the emitted spectrum we focus upon is the secondary peak in the luminosity at $t\sim 1.3\;{\rm s}$ seen in Figure (\ref{fig:MeVsIvo}). 
The spectrum as calculated for the portion of the signal after $t=1.0\;{\rm s}$ is shown in Figure (\ref{fig:Events3of3}) while in Figure (\ref{fig:EventsVsTime}) we observe the ability of Super-K, Hyper-K, JUNO, and DUNE to observe this feature as a function of time. Clearly this is a difficult feature to observe: in order to see the secondary DDT peak with Hyper-Kamiokande we find the supernova would need to be as close as 100 pc and even closer before the other three detectors become sensitive to the DDT peak.

At these late times we also found a peak around $E_{\nu} \sim 10\;{\rm MeV}$ appeared in the spectrum. This feature can be seen in Figure (\ref{fig:Events3of3}). The figure also indicates the 10-MeV peak we find at late times in the spectrum would only be observable in Hyper-K if the supernova were no further than around 10 pc. Observing this potentially distinguishing feature of the DDT model will require future multi-megaton class detectors and a  fortunately close supernova.

Figure (\ref{fig:Events3of3NoOscComp}) explains why the secondary peak in luminosity and spectral feature at $E_{\nu} \sim 10\;{\rm MeV}$ are so hard to observe. The figure plots the interaction event rate at Hyper-K and DUNE per 0.5 MeV for the signal after $t=1.0\;{\rm s}$. The dark lines represent the signal \emph{with} neutrino oscillation included while the lighter bands indicate the event rates when oscillations are removed.
As we have previously stated, oscillations decrease the expected event rates at all epochs and the decrease is particularly severe at late times. Oscillations convert an initial spectrum that is dominated by $\nu_e$ into a flux on Earth that is dominated by $\nu_\mu$ and $\nu_\tau$. This means that to capture more of the incoming flux, sensitivity to neutral-current processes needs to be increased. In particular, increasing sensitivity to neutral-current processes would show the biggest gain.


\subsubsection{Smearing}
\begin{figure}[ht]
	\centering
    \begin{minipage}{0.99\textwidth}
        \centering
        \includegraphics[trim={0 0 0 1.3cm},clip,width=1\linewidth]{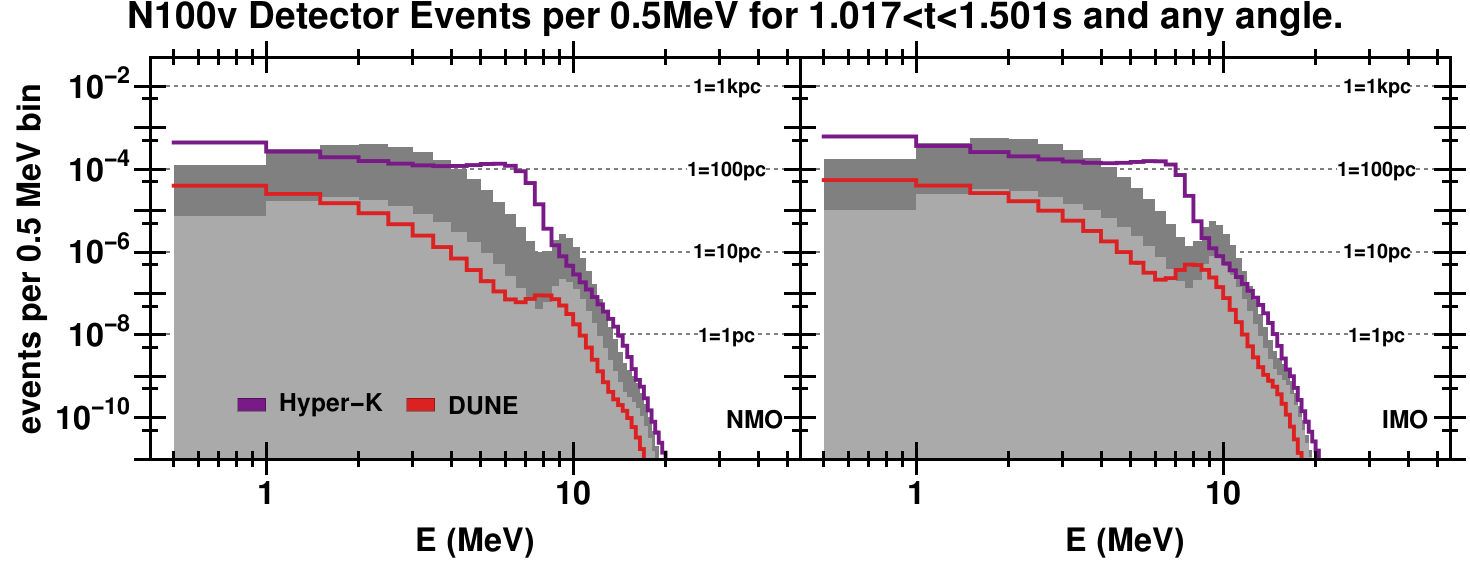}
    \end{minipage}
        \caption{Same as Figure (\ref{fig:Events3of3}) but with 0.5-MeV bins and smearing. The dark (light) gray is the Hyper-K (DUNE) event counts depicted in Figure (\ref{fig:Events3of3}). Note that the x-axis is now the measured energy.}
        \label{fig:Events3of3Smearing}
\end{figure}

Thus far, we have only presented interaction rates. A more complete prediction of experimental observations would need to include effects such as detector smearing and efficiencies (which include thresholds). The smearing effects are not yet fully determined for the detectors that are still in the design phase. Yet if reasonable estimates based on existing detectors of a similar type are made for the smearing, then SNOwGLoBES can be used to calculate the smeared rates.  Note that the smeared output from SNOwGLoBES accounts for distribution of interaction products as well as detector effects. Such a calculation (assuming 100\% post-smearing efficiency) reveals that the rates in Table (\ref{table:Events}) and the data presented in Figure (\ref{fig:EventsVsTime}) are essentially unchanged. The changes to Figure (\ref{fig:Events}) due to smearing  do not reveal anything unexpected. In contrast, the effects of smearing on Figure (\ref{fig:Events3of3}) are significant and are presented in Figure (\ref{fig:Events3of3Smearing}). This figure shows the smeared events for Hyper-K and DUNE together with gray unsmeared events. The dark (light) gray histograms represent the unsmeared Hyper-K (DUNE) rates. The comparison between the smeared and unsmeared results shows that Hyper-K can no longer distinguish the 10-MeV peak (a neutrino production feature). The smeared DUNE rates still show sensitivity to the 10-MeV peak for a sufficiently close SN, but smearing has shifted the peak to $\sim$8 MeV. 

These results are optimistic because detector efficiencies are not included and these would serve to lower observed rates. However it is also possible that future detectors will improve resolution.  For Figure (\ref{fig:Events3of3Smearing}), efficiencies are not expected to greatly alter a detector's ability to resolve a 10-MeV feature and so, for the purposes of observing the 10-MeV feature, the smeared results presented here reflect a conservative estimate of detection ability. The truth would lie somewhere between the smeared and unsmeared rates.

\subsubsection{IceCube}
Though not designed for low energies, the IceCube detector located in the ice of Antarctica also has sensitivity to the neutrinos from a supernova. 
However IceCube is different from the other four detectors considered because it does not have energy or directional resolution at MeV energies. The sensitivity to the neutrinos arises as an overall increase in the low-energy background rate of the detector. The large volume of IceCube and its excellent time resolution means that it may be possible for IceCube to detect energy integrated neutrino production or oscillation time-evolution features. The background rate in IceCube is not currently calculated by \textsc{SNOwGLoBES}, so in Appendix \ref{Appendix:IceCube} we detail how we calculate the rates.

   
\begin{figure}[t]
	\centering
    \begin{minipage}{0.95\textwidth}
        \centering
        \includegraphics[trim={0 0 0 1.3cm},clip,width=1\linewidth]{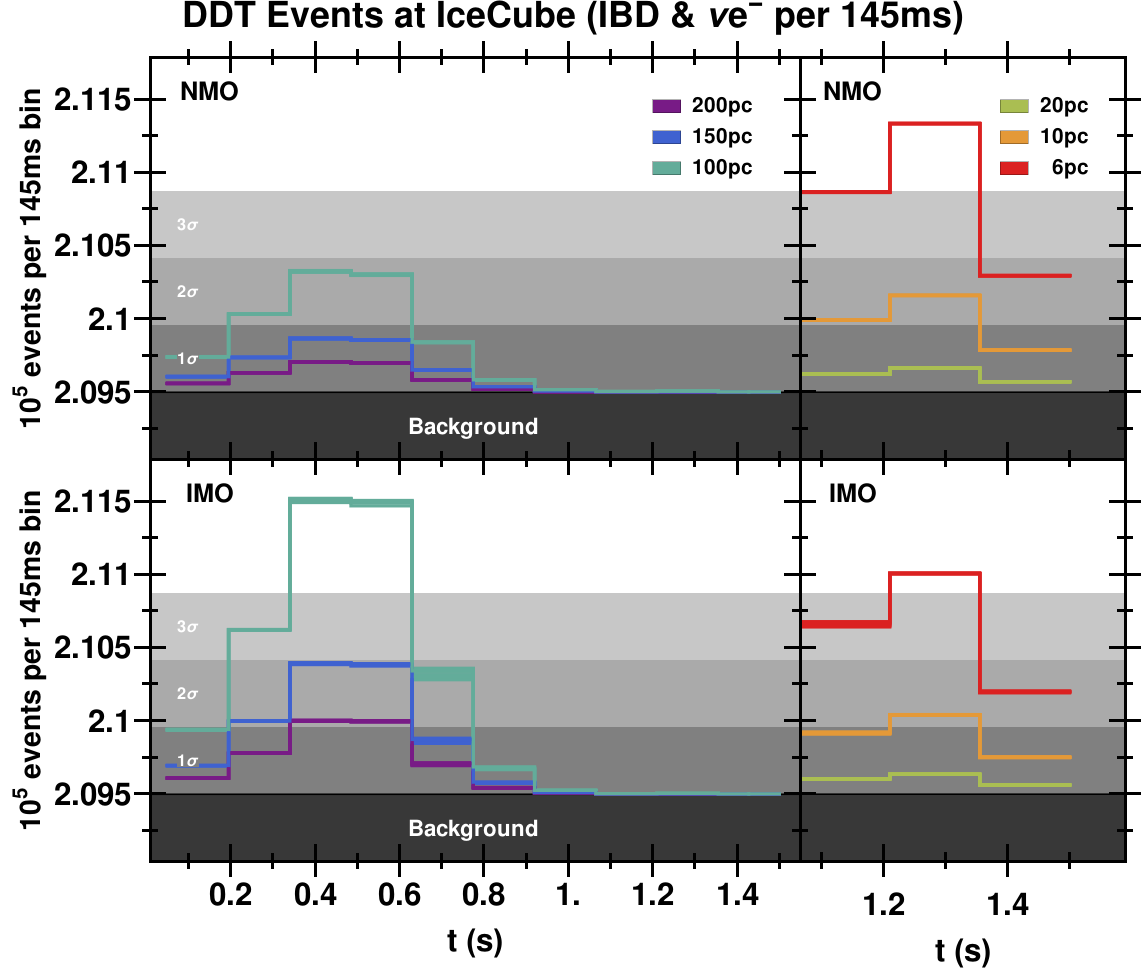}
    \end{minipage}
        \caption{
         SN Ia event counts in IceCube in 145 ms time bins. All eight trajectories are plotted and the gray shaded bands represent the background plus the denoted $\sigma$ to indicate how many events it would take to be perceived above a statistical fluctuation of the background. The top row is for normal mass ordering and the bottom row is for inverted mass ordering. The left column plots show the predicted number of events for the full 1.5 seconds of the neutrino signal and for sample SNe at 200, 150, and 100 pc. These distances were chosen to illustrate how close a SN would need to be in order for IceCube to observe it with statistical significance. The right column plots show the predicted number of events in the last three time bins where the sample SNe are now at 20, 10, and 6 pc. These distances were chosen to illustrate how close a SN would need to be in order for IceCube to detect the DDT at $\sim 1.3$1.3 s with statistical significance.}
        \label{fig:IceCubeEventsVsTime145}
\end{figure}

Table (\ref{table:Events}) shows that IceCube has an event rate a little below the interaction rate in Hyper-K and Figure (\ref{fig:IceCubeEventsVsTime145}) shows the predicted number of events above the background from the inverse beta decay (IBD) and electron scattering events in IceCube using 145 ms time bins. The top plots are for a NMO, the bottom are for an IMO. The left plots show the full 1.5s neutrino signal and the right plots show just the final 3 time bins (but for different example SN distances than those used in the left plots). This figure reveals that one would need a N100$\nu$ SN at within 200 pc to see a 1$\sigma$ deviation at the emission peak.  This figure also reveals that one would need a N100$\nu$ SN at approximately 10 pc to see a $\sim1\sigma$ detection of the DDT at $\sim 1.3$s. There are almost no SN Ia candidates within 10 pc and thus it is unlikely that a type Ia Supernova will occur close enough for IceCube to observe the DDT. Even with a SN as near as 200 pc, the maximum signal in IceCube is statistically weak and it will fall to future upgrades of IceCube to attempt the detection of type Ia supernovae.


\section{Conclusion \label{sec:Conclusion}}

Type Ia supernovae do emit neutrinos and there is information about the explosion in the signal. 
In this paper we computed the neutrino signal as a function of energy and time from a DDT simulation in a variety of different detectors. 
In addition to the secondary peak in luminosity at $t\sim 1.3\;{\rm s}$ noted by Seitenzahl \emph{et al.} \cite{Seitenzahl2015a}, we found that a spectral peak at $E\sim10\;{\rm MeV}$ emerges in the spectrum at these late times due to electron capture on copper. 
This peak is a potentially distinguishing feature of explosion models since it reflects the nucleosynthesis conditions early in the explosion. 

We computed the full energy and time dependence of the transition probability through the supernova using a three-flavor evolution code along eight representative lines of sight through the supernova. 
We found time dependence in the transition probabilities as a function of energy for every 
line of sight considered and a variation between lines of sight emerged between $t\sim 0.8\;{\rm s}$ and $t\sim 1.3\;{\rm s}$. Along each line of sight the general trend is for the neutrino oscillations to become more adiabatic as time progresses. 
At late times the electron neutrino flux on Earth is approximately an order of magnitude smaller in the normal mass ordering due to oscillations. 

When we computed the event rates in the largest current and next-generation neutrino detectors we found a type Ia supernova at the most probable distance of $d\sim10\;{\rm kpc}$ will barely be visible. 
Distinguishing between different near-Chandrasekhar mass explosion models, which all yield similar order-of-magnitude neutrino luminosities (compare \cite{kunugise2007a}, \cite{Odrzywolek2011a}, and this work), is therefore not very promising. We note, however, that detecting any neutrinos from a SN Ia at all would be a strong indication of explosive nuclear burning at densities above $10^9\;\mathrm{g cm^{-3}}$ and hence a clear sign of a deflagration in a near-Chandrasekhar mass WD. The argument why we could exclude popular models involving only detonations in less massive WDs, such as the violent merger models \cite{pakmor2013a} or the double detonation models \cite{fink2010a}, is simple. The neutrino luminosity of NSE material, which we demonstrated is the dominant channel, is at least 2 orders of magnitude lower for the central densities ($<10^8\;\mathrm{g cm^{-3}}$) of these models, see figure 16 from \cite{calder2007a}.

Not only is the overall luminosity of the supernova significantly smaller than for core-collapse supernovae, the spectrum peaks at energies of order $E\sim 1\;{\rm MeV}$ rather than the $E\sim 10-20\;{\rm MeV}$ for core-collapse supernovae. 
For a close type Ia supernova, $d = 1\;{\rm kpc}$, JUNO, Super-K, and DUNE will record a few events while Hyper-K will observe several tens of events. 
If statistics permit partitioning of the signal, the secondary peak in luminosity at $t\sim 1.3\;{\rm s}$ and the spectral feature at $E_{\nu} \sim 10\;{\rm MeV}$ will be difficult to observe because of the increased adiabaticity of the neutrino evolution at this epoch. IceCube has a SN Ia detection sensitivity comparable to that of DUNE, JUNO, and Super-K but needs a much closer SN Ia in order to observe the second luminosity peak.
   

\section{Acknowledgments}
We are grateful to Evan O'Connor for his help with the implementation of \textsc{NuLib}.
This work was supported at NC State by DOE grant DE-SC0006417. Seitenzahl acknowledges support from the Australian Research Council Laureate Grant FL0992131. Scholberg's research activities are supported by the U.S. Department of Energy and the National Science Foundation.


\appendix
\section{IceCube Event Rate\label{Appendix:IceCube}}
The event rate at IceCube is calculated by summing the electron elastic scattering (ES) event rate for each of the 6 neutrino species together with the inverse beta decay (IBD) rate. Following \cite{Dighe2003}, the number of useful Cherenkov photons released by a moving electron or positron is $N_\gamma=191E_{e^\pm} \text{ MeV}^{-1}$ and the target density is
\begin{align}
    \rho_T\left(n_T\right)&=n_T\left(0.924\frac{\text{g}}{\text{cm}^3}\right)
    	\frac{N_A}{\text{mol}}\left(\frac{\text{mol}}{18.01528\text{ g}}\right),
\end{align}
where $N_A$ is Avogadro's number and $n_T$ represents the number of targets per water molecule ($n_T=10$ for electrons and $n_T=2$ for protons). The electron \cite{Giunti-Kim-2007} or positron \cite{2011IceCubeSNe} energy can be derived in terms of the neutrino energy as
\begin{align}
	E_{e^+,IBD}\left(E_{\bar{\nu}_e}\right)&=\left(E_{\bar{\nu}_e}-\frac{m_N^2-m_P^2-m_e^2}{2m_P}\right)\left(1-\frac{E_{\bar{\nu}_e}}{E_{\bar{\nu}_e}+m_P}\right),\\
	E_{e^-,ES}\left(E_\nu,c_\theta\right)&=m_e\frac{\left(m_e+E_\nu\right)^2+E_\nu^2c_\theta^2}{\left(m_e+E_\nu\right)^2-E_\nu^2c_\theta^2},
\end{align}
where $c_\theta=\cos\theta$ and $\theta$ is the electron recoil angle. Also, $E_{e^-,ES}$ is set to zero if it is less than the Cherenkov threshold of 0.79 MeV which effectivly sets $N_\gamma=0$ for sub-threshold electrons.  The cross section \cite{Giunti-Kim-2007} for IBD is $\sigma_{IBD}=9.52\times10^{-44}E_{\bar{\nu}_e}$ and for ES it is
\begin{align}
	\frac{\text{d}\sigma_{ES}}{\text{d}c_\theta}&=\sigma_0
    	\frac{4E_\nu^2c_\theta\left(m_e+E_\nu\right)^2}{\left(\left(m_e+E_\nu\right)^2-E_\nu^2c_\theta^2\right)^2}
        \left(
        	g_1^2+g_2^2\left(1-\frac{2m_e E_\nu c_\theta^2}{\left(m_e+E_\nu\right)^2-E_\nu^2c_\theta^2}\right)^2
            -g_1 g_2\frac{2m_e^2  c_\theta^2}{\left(m_e+E_\nu\right)^2-E_\nu^2c_\theta^2}
        \right),
\end{align}
with $\sigma_0 =8.8059\times 10^{-45}\text{cm}^{2}$ and $g_1$ and $g_2$ defined as
\begin{align}
    \begin{pmatrix} g_1 \\ g_2 \end{pmatrix} &=
    	\begin{pmatrix} 0 & 1 \\ 1 & 0 \end{pmatrix}^\eta \begin{pmatrix} \sin^2\theta_\text{W} \\ \pm0.5+\sin^2\theta_\text{W} \end{pmatrix},
\end{align}
where $\eta$ is 0 for antineutrinos and 1 for neutrinos, $+0.5$ is for electron neutrinos and $-0.5$ is for muon or tau neutrinos, and $\sin^2\theta_\text{W}\approx0.23$ is the Weinberg angle. The current configuration of IceCube has 5160 Optical Modules (OM), each of which has an effective volume of about $V_{eff}=190600\text{ cm}^3$ \cite{Dighe2003,Halzen2009}. The last ingredient is the oscillated flux measured on Earth from a DDT SN Ia at 10kpc, $\Phi\left(E_\nu,t\right)$ (calculated in Section \S\ref{subsec:OcsResults}). Putting it all together yields

\begin{align}
	N_{ES}\left(D\right)&=5160\left(\frac{10^4}{D}\right)^{2}
      \int_{t_i}^{t_f}\int_{E_{\nu,i}}^{E_{\nu,f}}\int_0^1\left(
          \rho_T\left(10\right) V_{eff} \Phi\left(E_\nu,t\right)\frac{\text{d}\sigma_{ES}}{\text{d}c_\theta}\left(E_\nu,c_\theta\right)
               N_{\gamma}\left(E_\nu,c_\theta\right)
      \right)\text{d}c_\theta\text{d}E_\nu\text{d}t\\    
	N_{IBD}\left(D\right)&=5160\left(\frac{10^4}{D}\right)^{2}
      \int_{t_i}^{t_f}\int_{E_{\bar{\nu}_e,i}}^{E_{\bar{\nu}_e,f}}\left(
          \rho_T\left(2\right) V_{eff} \Phi\left(E_{\bar{\nu}_e},t\right)\sigma_{IBD}\left(E_{\bar{\nu}_e}\right)
               N_{\gamma}\left(E_{\bar{\nu}_e}\right)
      \right)\text{d}E_\nu\text{d}t
\end{align}
where D is the SN distance in parsecs. These event rates are what are plotted in Figure (\ref{fig:IceCubeEventsVsTime145}) and  need to be compared to the background rate of $280\text{ s}^{-1}$ in each OM \cite{Halzen2009}.

\section{Event Spectrum Channel Breakdown\label{Appendix:Channel}}
\begin{figure}[t]
	\centering
    \begin{minipage}{0.95\textwidth}
        \centering
        \includegraphics[trim={0 0 0 0},clip,width=1\linewidth]{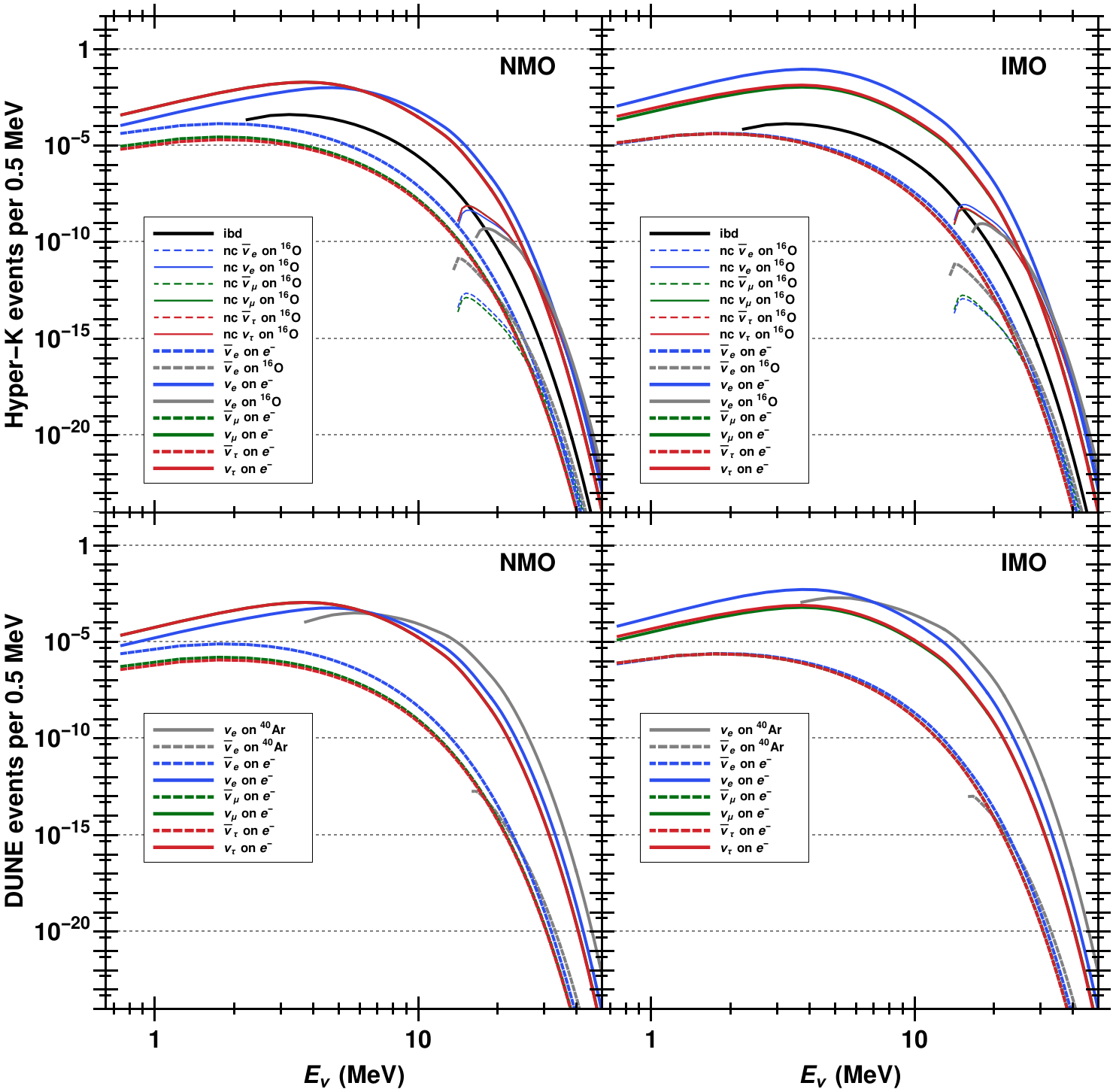}
    \end{minipage}
        \caption{SN Ia interaction event counts per 0.5 MeV per interaction channel in Hyper-K and DUNE. The left plots are for normal mass ordering and the right plots are for inverted mass ordering. The top plots are for Hyper-K and the bottom ones are for DUNE. The event counts are for the full $\sim 1.5$ s neutrino signal.}
        \label{fig:SpectrumChannelBreakdown}
\end{figure}

\begin{figure}[t]
	\centering
    \begin{minipage}{0.95\textwidth}
        \centering
        \includegraphics[trim={0 0 0 0},clip,width=1\linewidth]{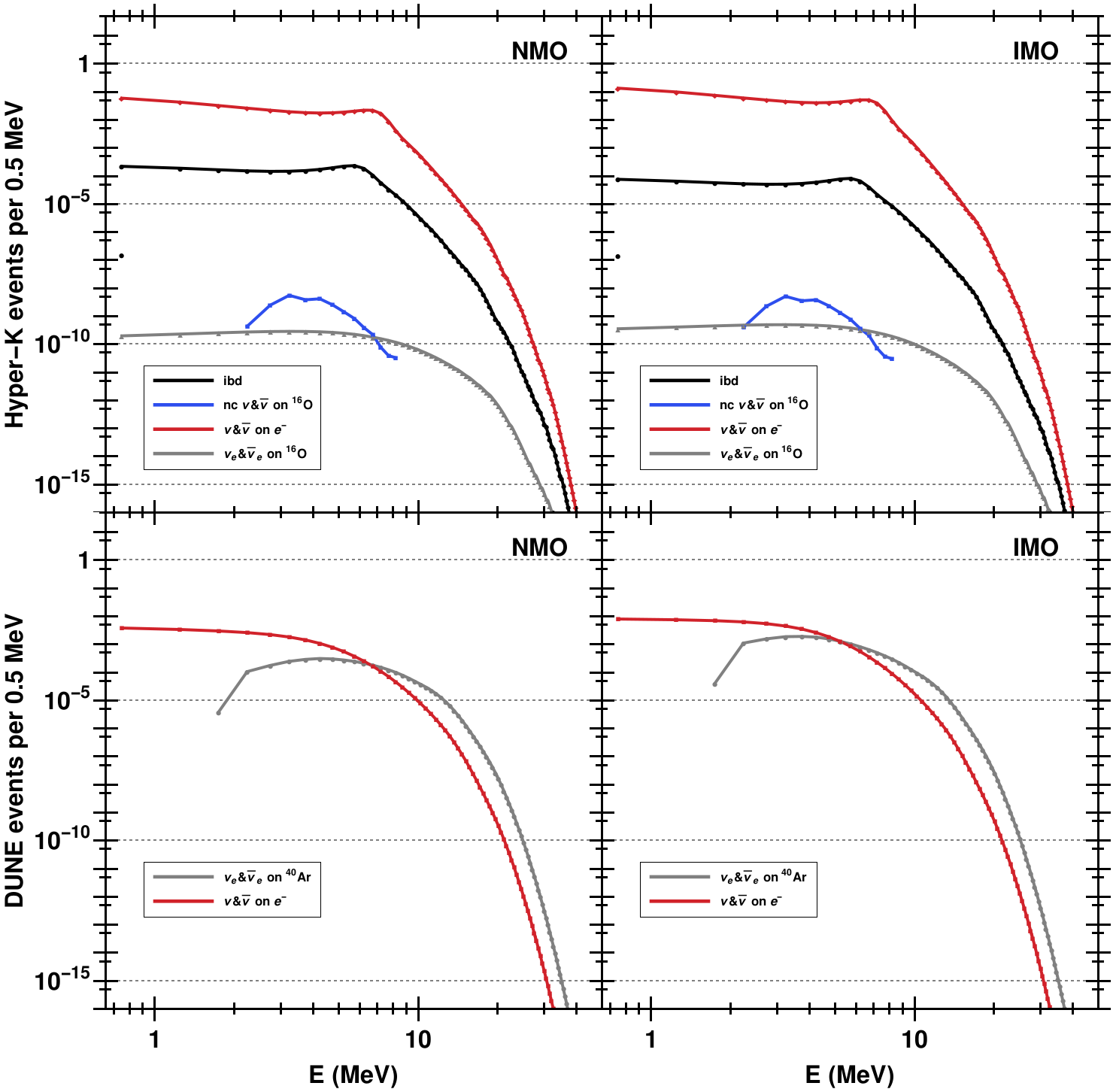}
    \end{minipage}
        \caption{SN Ia event counts per 0.5 MeV per interaction channel group in Hyper-K and DUNE. The left plots are for normal mass ordering and the right plots are for inverted mass ordering. The top plots are for Hyper-K and the bottom ones are for DUNE. The event counts are for the full $\sim 1.5$s neutrino signal and include energy smearing from the neutrino energy to the measured particle's energy (x-axis).}
        \label{fig:SpectrumChannelGrpBreakdown}
\end{figure}

In this section, Figure (\ref{fig:SpectrumChannelBreakdown}) is presented in order to show the individual interaction channel contributions to the interaction event rates in the Hyper-K and DUNE detectors. These individual rates may be used to determine the most effective search strategy. The interaction event rate is per 0.5-MeV  bin and is for the full $\sim 1.5$ s neutrino signal. From the top Hyper-K plots it is clear that the neutrino interactions (all flavors) with free electrons are the dominant contributions for both mass orderings, even above IBD. For DUNE (bottom plots), Figure (\ref{fig:SpectrumChannelBreakdown}) shows that the neutrino interactions (all flavors) with free electrons are about as important as electron neutrino interactions with $^{40}$Ar. However, the anti-neutrino interactions (all flavors) with free electrons are much more important than the anti-electron neutrino interactions with $^{40}$Ar (for the overall rate).  Figure (\ref{fig:SpectrumChannelGrpBreakdown}) is similar to Figure (\ref{fig:SpectrumChannelBreakdown}) except that the event counts are per interaction channel group and now approximately account for the energy smearing from the neutrino energy to the measured particle's energy. This figure shows which channel groups will produce the greatest number of detector events.

\bibliographystyle{IEEEtran}
\bibliography{main}

\end{document}